\documentclass[reprint,amsmath,amssymb,aps,prb,superscriptaddress,showpacs,citeautoscript,floatfix]{revtex4-1}
\usepackage{graphicx}
\usepackage{color}
\usepackage{dcolumn}
\usepackage{bm}
\usepackage{amsmath}
\usepackage{amssymb}
\usepackage[]{units}
\usepackage{subcaption}
\usepackage{ragged2e}
\DeclareCaptionJustification{justified}{\justifying}
\usepackage[justification=justified, format=plain]{caption}
\usepackage{booktabs}
\DeclareGraphicsExtensions{.eps,.png,.jpg,.pdf}
\graphicspath{{Figures/}}

\begin{document}

\title[Phase stability of {Fe}{Rh}]{Impact of lattice dynamics on the phase stability of
metamagnetic FeRh: Bulk and thin films}

\author{M. Wolloch}
\email{mwo@cms.tuwien.ac.at}
\affiliation{Institute of Applied Physics, Vienna University of Technology, Wiedner Hauptstr. 8-10/134, 1040 Vienna, Austria}
\author{M. E. Gruner}
\affiliation{Faculty of Physics and Center for Nanointegration Duisburg-Essen (CENIDE), University of Duisburg-Essen, 47048 Duisburg, Germany}
\author{W. Keune}
\affiliation{Faculty of Physics and Center for Nanointegration Duisburg-Essen (CENIDE), University of Duisburg-Essen, 47048 Duisburg, Germany}
\author{P. Mohn}
\affiliation{Institute of Applied Physics, Vienna University of Technology, Wiedner Hauptstr. 8-10/134, 1040 Vienna, Austria}
\author{J. Redinger}
\affiliation{Institute of Applied Physics, Vienna University of Technology, Wiedner Hauptstr. 8-10/134, 1040 Vienna, Austria}
\author{F. Hofer}
\affiliation{Institute of Solid State Physics, Vienna University of Technology, Wiedner Hauptstr. 8-10/134, 1040 Vienna, Austria}
\author{D. Suess}
\affiliation{Institute of Solid State Physics, Vienna University of Technology, Wiedner Hauptstr. 8-10/134, 1040 Vienna, Austria}
\author{R. Podloucky}
\affiliation{Institute of Physical Chemistry, University of Vienna , Sensengasse 8/7 A-1090 Vienna, Austria}
\author{J. Landers}
\affiliation{Faculty of Physics and Center for Nanointegration Duisburg-Essen (CENIDE), University of Duisburg-Essen, 47048 Duisburg, Germany}
\author{S. Salamon}
\affiliation{Faculty of Physics and Center for Nanointegration Duisburg-Essen (CENIDE), University of Duisburg-Essen, 47048 Duisburg, Germany}
\author{F. Scheibel}
\affiliation{Faculty of Physics and Center for Nanointegration Duisburg-Essen (CENIDE), University of Duisburg-Essen, 47048 Duisburg, Germany}
\author{D. Spoddig}
\affiliation{Faculty of Physics and Center for Nanointegration Duisburg-Essen (CENIDE), University of Duisburg-Essen, 47048 Duisburg, Germany}
\author{R. Witte}
\affiliation{Institute of Nanotechnology, Karlsruhe Institute of Technology, 76344 Eggenstein-Leopoldshafen, Germany}
\author{B. Roldan Cuenya}
\affiliation{Department of Physics, Ruhr-University Bochum, 44780 Bochum, Germany}
\author{O. Gutfleisch}
\affiliation{Materials Science, TU Darmstadt, 64287 Darmstadt, Germany}
\author{M. Y. Hu}
\affiliation{Advanced Photon Source, Argonne National Laboratory, Argonne, Illinois, 60439, USA}
\author{J. Zhao}
\affiliation{Advanced Photon Source, Argonne National Laboratory, Argonne, Illinois, 60439, USA}
\author{T. Toellner}
\affiliation{Advanced Photon Source, Argonne National Laboratory, Argonne, Illinois, 60439, USA}
\author{E. E. Alp}
\affiliation{Advanced Photon Source, Argonne National Laboratory, Argonne, Illinois, 60439, USA}
\author{M. Siewert}
\affiliation{Faculty of Physics and Center for Nanointegration Duisburg-Essen (CENIDE), University of Duisburg-Essen, 47048 Duisburg, Germany}
\author{P. Entel}
\affiliation{Faculty of Physics and Center for Nanointegration Duisburg-Essen (CENIDE), University of Duisburg-Essen, 47048 Duisburg, Germany}
\author{R. Pentcheva}
\affiliation{Faculty of Physics and Center for Nanointegration Duisburg-Essen (CENIDE), University of Duisburg-Essen, 47048 Duisburg, Germany}
\author{H. Wende}
\affiliation{Faculty of Physics and Center for Nanointegration Duisburg-Essen (CENIDE), University of Duisburg-Essen, 47048 Duisburg, Germany}

\begin{abstract}
We present phonon dispersions, element-resolved vibrational density of states (VDOS)
and corresponding thermodynamic properties
obtained by a combination of density functional theory (DFT) and nuclear
resonant inelastic X-ray scattering (NRIXS) across the metamagnetic transition of B2 FeRh
in the bulk material and thin epitaxial films.
We see distinct differences
in the VDOS of the antiferromagnetic (AF) and ferromagnetic (FM) phase
which provide a microscopic proof of strong spin-phonon coupling in FeRh.
The FM VDOS exhibits a particular sensitivity to the slight tetragonal
distortions present in epitaxial films, which is not encountered in the AF phase.
This results in a notable change in lattice entropy,
which is important for the comparison between
thin film and bulk results.
Our calculations confirm the recently reported lattice instability
in the AF phase. The imaginary frequencies at the $X$-point depend critically on
the Fe magnetic moment and atomic volume. Analyzing these non vibrational modes leads to the discovery of a stable monoclinic ground state structure which is robustly predicted from DFT but
not verified in our thin film experiments.
Specific heat, entropy and free energy calculated within the quasiharmonic 
approximation suggest that the new phase is possibly suppressed because of its relatively smaller lattice entropy. 
In the bulk phase, lattice degrees of freedom contribute with the same sign and
in similar magnitude to the isostructural AF-FM phase transition as the
electronic and magnetic subsystems and therefore needs to be included in thermodynamic modeling.

\end{abstract}


\maketitle

\section{Introduction}
\label{sec:Intro}
During recent years, ordered B2 FeRh (CsCl structure) has received increased attention due to its extraordinary properties,
in particular its temperature-driven
isostructural transition between a ferromagnetic (FM) and antiferromagnetic (AF)
phase at $T_{\rm M}\sim\unit[350]{K}$, which
was discovered more than seven decades ago~\cite{fallot:38,fallot:39,shirane:63,shirane:64}.
This transition is accompanied by a large volume change of $\sim 1\,$\% and a
complete loss of the Rh moment in the AF phase (1.0\,$\mu_{\rm B}$ in FM),
while the Fe moment remains large and essentially constant around 3.2-3.3\,$\mu_{\rm B}$
across the transition.
The availability of a metamagnetic transition near room temperature (RT)
makes FeRh an interesting material for a number of technological applications like
heat assisted magnetic recording (HAMR)\cite{weller:14}, antiferromagnetic spintronics\cite{jungwirth:16} and magnetic refrigeration~\cite{yu:10}. For a recent review on this topic see Ref.~\onlinecite{lewis:16}.

HAMR is believed to be the future magnetic recording technology
in order to extend the areal density to \unit[4]{Tb/in$^2$} and beyond~\cite{weller:14}. Near field transducers (NFC) are used to focus laser light well below the diffraction limit to give a thermal write assist, enabling the use of highly anisotropic recording media like FePt~\cite{stipe:10,challener:09}.
Due to the high Curie temperature of FePt, thermally written-in-errors due to highly excited states in combination with low saturation magnetization, as well as the limited lifetime of NFCs remain an issue for this technology~\cite{richter:12,budaev:12}.
An interesting idea to overcome those problems is to replace the second order transition of FePt by the first order transition of FeRh. Thiele et al.\ proposed an exchange spring structure for the recording process by coupling FeRh to FePt~\cite{thiele:03}. The advantage is that the first order phase transition of FeRh can be tailored well below the Curie temperature of FePt, which relaxes the lifetime and reliability problem of the NFCs. Furthermore during recording the magnetic moment of FeRh is still high overcoming the problem of thermally written-in-errors \cite{suess:15}.

Another exciting application of FeRh is in the developing field of spintronics, which is promising significant advantages in data storage~\cite{chappert:07}. It has been shown that the AF to FM transition in FeRh can be driven by electric fields, leading to electric on and off switching of ferromagnetism near RT~\cite{cherifi:14}. In contrast to FM memory and storage devices, application based on AF spintronics are insensitive to magnetic field pertubation and generate no magnetic stray fields, thus eliminating crosstalk. These advantages come at the price of an increased difficulty in manipulating the antiferromagnetic aligned magnetic moments~\cite{gomonay:14,jungwirth:16}.
A room temperature FeRh AF memory resistor was recently demonstrated by Marti and coworkers, using the FM phase to pre-align the magnetic moments of the AF phase after cooling, and using the anisotropic magnetoresistance (AMR) to read out data~\cite{marti:14}.
Soon after, sequential write-read operations on FeRh AF memory were performed using
Joule heating to trigger the metamagnetic transition~\cite{matsuzaki:15,moriyama:15}.
Employing M\"ossbauer spectroscopy, Bordel et al.\ reported a strain-driven Fe-spin reorientation across the AF-to-FM transition in epitaxially strained FeRh thin films grown on MgO, which could be of significant use in AF spintronics applications~\cite{bordel:12}. It has also been shown that FeRh can be used as a material for spin wave transmission~\cite{usami:16}.

Recently, also the giant magnetocaloric effect (MCE) of B2 FeRh
shifted back into the scientific focus, since
FeRh has one of the highest adiabatic temperature changes of all known
materials,~\cite{annaorazov:92,gschneidner:00,manekar:08,liu:12,chirkova:16,zverev:16}
which in addition is accompanied by large elasto- and barocaloric
effects~\cite{nikitin:92,stern-taulats:14,stern-taulats:15}.
However, the transition suffers from a large hysteresis and a high sensitivity to
stoichiometry and anti-site defects.\cite{staunton:14}
Nonetheless, Liu et al.\ reported a large reversible caloric effect
in a dual-stimulus magnetic-electric refrigeration cycle
for a FeRh film grown epitaxially on BaTiO$_3$~\cite{liu:16}.

The origin of the metamagnetic transition has been under
vibrant debate for more than half a century.
An early explanation was given using the exchange inversion model by Kittel, where the exchange parameters vary linearly with the lattice parameter and change their sign at some critical value, thus causing the transition~\cite{kittel:60}. However, this model is incompatible with
the large entropy change observed at the
transition~\cite{kouvel:66,lommel:69,mckinnon:70,annaorazov:96}.
Later, reasoned by the large differences in low temperature specific heats between the AF and FM phases, Tu et al.\cite{tu:69} argued that a change in entropy of band electrons is
solely responsible for the transition, but this view fails to explain
the transition in the case of admixture of 5\% Ir to FeRh,
where the relation of the specific heats becomes reversed~\cite{kouvel:66,ivarsson:71,fogarassy:72}.

In 2003, part of the present authors proposed an explanation
for the transition based on the competition between AF Fe-Fe and FM Fe-Rh
exchange interactions. Monte Carlo simulations of a Blume-Capel spin model suggested 
that longitudinal thermal fluctuations of the Rh magnetic moments in the FM phase
give rise to a Schottky anomaly far below $T_{\rm M}$, which finally provides
the entropic stabilization of the FM phase.\cite{gruner:03}
Coupling the magnetic subsystem to lattice and volume degrees of freedom
leads to a very good agreement with experiment
in terms of the entropy change $\Delta S$ at $T_{\rm M}$ and the temperature dependence
of the Gibbs free energy difference $\Delta G$.
Recent state-of-the-art measurements
of the specific heat $C_p$ in FeRh thin films with AF and FM magnetic order seem to
support the presence of this Schottky anomaly.\cite{cooke:12}
However, Cooke et al.\ separated the magnetic contribution from the other degrees of freedom
in terms of a simple Debye model due to the lack of reliable lattice vibration data.
Based on their analysis these authors proposed a strong competition between
large magnetic and lattice contributions to $\Delta S$ with opposite sign.

In 2004, Mryasov showed that it is also
possible to model the transition with a Heisenberg
Hamiltonian including bi-quadratic exchange interactions.\cite{ju:04,mryasov:04}
In contrast to Ref.\ \onlinecite{gruner:03}, the Rh moments were treated as
induced by the magnetic moments of the surrounding Fe.
Derlet introduced an empirical Landau-Heisenberg model with parameters fitted to existing ab-initio calculations. The paper concluded that a quadratic exchange term is needed to produce the transition and that both volume- and magnetic fluctuations are equally important~\cite{derlet:12}.
Barker and Chantrell extended Mryasovs model by fully expanding the
quadratic spin interactions into four spin exchange terms, which were 
parameterized from experimental data.\cite{barker:15}
Solving the Landau-Lifshitz-Gilbert equation using
atomistic spin dynamics yields $T_{\rm M}$ in good agreement with experiment.

The DFT calculations of Sandratskii and Mavropoulos pointed out that
in the AF phase hybridization between Fe and Rh states
causes a local spin polarization of Rh, which averages to zero.\cite{sandratskii:11}
Later on, Kudrnovsk\'y, Drchal, and Turek argued that the
hybridization with surrounding Fe moments is the main
reason for the development of magnetic moments on Rh atoms
in the FM phase.\cite{kudrnovsky:15}
The importance of hybridization effects in both, AF and FM speaks
against a simple Stoner-like picture of an induced Rh moment
and rather for the presence of metastable magnetic states of Rh.

Recent work also aimed at incorporating finite temperature changes
to the electronic structure arising from magnetic excitations.
De\`ak and coworkers evaluated the magnetic and electronic contributions to $\Delta G$ 
in their relativistic disordered local moment (DLM) approach.\cite{deak:14}
They were able to reproduce a transition from AF to FM, albeit at
a rather large temperature and atomic volume.
In a similar spirit Polesya et al.\ evaluated an extended Heisenberg Hamiltonian
which only acts on the Fe atoms, where the FM exchange interaction
is scaled according to a response function dependent on the Rh moments.\cite{polesya:16}
With magnetic exchange parameters self-consistently
obtained from uncompensated DLM calculations,
which correspond to the average moment of the configuration, they
obtain a very reasonable transition temperature of \unit[$\sim320$]{K}.

A completely different route was taken by
Gu and Antropov, who derived the magnetic contributions to the free energy and $\Delta S$
from the magnon (spin wave) spectra
calculated from first-principles.\cite{gu:05}
In this approach, neither Stoner-type nor spin-flip-excitations are considered.
From the comparison with experimental data, the authors concluded that the
magnetic degrees of freedom provide the dominant
contribution to the transition.

The relevance of a specific degree of freedom for the metamagnetic
transition is reflected in its temperature dependent
contribution to the specific heat $C_p$ and
finally also to $\Delta S$ and $\Delta G$. A direct comparison of these calculated quantities
with experimental data is thus inevitable to evaluate a specific model.
This requires
the detailed knowledge of all individual contributions to the Gibbs
free energy difference between the AF and FM phase $\Delta G(T,p)$,
which is usually divided into vibrational, magnetic and electronic degrees of freedom ($\Delta G=\Delta G^{\rm vib}+\Delta G^{\rm mag}+\Delta G^{\rm el}$).
So far, respective computational data are only
provided by Refs.~\onlinecite{gruner:03,gu:05,deak:14}. Experimental information on $C_p(T)$ is
available from  Refs.~\onlinecite{richardson:73, cooke:12}, while $\Delta G(T)$ was measured by Ponomarev.\cite{ponomarev:72}

This work aims at providing for the first time a comprehensive overview
of the lattice dynamical contribution to the phase stability
in FeRh, from both the computational as well as the experimental point of view.
With the notable exception of a pioneering inelastic neutron spectroscopy study
of the phonon dispersion along the [111] crystallographic direction in the AF state at room temperature~\cite{castets:77}, and the recent determination of the element-specific Debye-Waller-factor by X-ray spectroscopy~\cite{wakisaka:15}, no experimental study on lattice vibrations of B2 FeRh has been published to the best of our knowledge.
Very recently, a computational study reported unstable lattice dynamics in the B2 AF structure
in combination with a strain-induced softening of the shear elastic constant $C^{\prime}$.\cite{aschauer:16}
However, thermodynamic properties associated with the lattice degrees of freedom are still not available, yet.

After an introduction to methodological details (Sec.~\ref{sec:Method}),
we present in a thorough characterization of
electronic structure and phonon dispersion relations (Sec.~\ref{sec:structure}).
The latter imply an instability of the B2 AF phase
towards a new monoclinic or orthorhombic
low energy phase, which depends sensitively on the magnitude of the Fe moment.
In Sec.~\ref{sec:experiment} we compare the vibrational density of states (VDOS)
with results of $^{57}$Fe nuclear resonant inelastic X-ray scattering (NRIXS),
which deliver the Fe-projected VDOS of B2-ordered AF and FM FeRh thin films. From the latter, we derive Fe-projected thermodynamic quantities.
Finally in section~\ref{sec:thermo}, we evaluate from first-principles
the vibrational and electronical
contributions $\Delta G^\mathrm{vib}(T)$ and $\Delta G^\mathrm{el}(T)$ to the metamagnetic transition and for the proposed new low temperature phase
in the quasi-harmonic approximation.
The calculated results are found to be in excellent agreement with
available experimental data.

\section{Methodological Details}
\label{sec:Method}

\subsection{Electronic structure calculations}
\label{sub:Comp}
Spin polarized DFT computations have been performed
employing the Vienna Ab-Initio Simulation Package
\textit{VASP}~\cite{kresse1993,kresse1994a,kresse1996a,kresse1996b}
version 5.4.1, using the Projector Augmented-Wave (PAW)
method~\cite{bloechl1994,kresse:98}. A dense $\Gamma$-centered k-mesh
of $17\times17\times17$ points was used to sample the Brillouin zone
of our 4 atom fcc-like unit cell. Meshes of equal or greater density
were used for larger supercells. The plane wave energy cutoff was
chosen to be \unit[450]{eV}, more than 160\% (180\%) of the standard
value for the Fe (Rh) PAW potential (set of 2003) which treats the
$3d$ and $4s$ ($4p$, $5s$, and $4d$) electrons as valence. With this k-mesh and energy cutoff total energies are converged to less than
\unit[1]{meV} per formula unit. We employ different functionals to
describe the effects of exchange and correlation, to study the
functional dependence of our results. If not indicated otherwise the
Perdew-Burke-Ernzerhof version of the generalized gradient correction
(GGA) (PBE~\cite{PBE}) has been used. In addition we used the revised
Perdew-Burke-Ernzerhof (RPBE~\cite{RPBE}), the Perdew-Burke-Ernzerhof
revised for solids (PBEsol~\cite{PBEsol}), and the Perdew-Wang 91
(PW91~\cite{PW91}) GGAs. Moreover the van der Waals corrected
optB86b~\cite{optB86b} and the local density approximation
(LDA~\cite{LDA}) functionals have been employed in parts of this
work. To ensure accurate forces during relaxations and phonon
calculations we use an additional superfine fast Fourier transform
(FFT) grid for the evaluation of the augmentation charges and a
smearing of \unit[$\leq0.1$]{eV} according to Methfessel and
Paxton~\cite{methfessel:89} (first order). For total energy
calculations the tetrahedron method with Bl\"ochl corrections has been
used~\cite{bloechl:94}. In all total energy GGA calculations we explicitly account
for non spherical contributions of the gradient corrections inside the
PAW spheres.  Phonon calculations were carried out in the harmonic
approximation using both the small displacement method (usually \unit[0.01]{\AA}) and density
functional perturbation theory (DFPT), using the \textit{phonopy}~\cite{togo:08} and
\textit{PHON}~\cite{alfe:09} codes.

For the calculation of the thermodynamic contributions from the vibrational and electronic degrees of freedom in the
quasiharmonic approximation we employ a similar but slightly different setup.
We used valence states of $3p$, $3d$, and $4s$ for Fe and $4$, $4d$, and $5s$ for Rh with a plane wave
cutoff $E_{\rm cut}=450$\,eV. 
We used a $3 \times 3 \times 3$  supercell and mostly single displacements of about \unit[0.02]{\AA} 
to keep the numerical effort tractable.
Forces were determined using a
Monkhorst-Pack k-grid of $4 \times 4 \times 4$ (except for the orthorhombic $Pmm$2
structure, where, according to the larger primitive cell and the thoroughly stable
phonon dispersion, we reduced the k-mesh to $2 \times 2 \times 2$) in combination
with a finite temperature smearing according to Methfessel and
Paxton\cite{methfessel:89}
with a broadening of $\sigma=0.1$\,eV.
For the calculations of the thermodynamic properties from the vibrational density of states,
we could safely neglect the imaginary modes in the B2(AF) phase. These occur only at small
lattice constants $a_0 \leq 3.02\,$\AA\ and are present only in a very small fraction of the reciprocal space; even when imaginary modes were omitted, the integrated density of states deviates from unity by less than
$0.1\,$\%.
Consequently, a comparison of the thermodynamic quantities with the results of
computationally much more demanding calculations for fully relaxed configurations
did not result in a notable difference.
The thermodynamic quantities of the electronic subsystems were calculated in a similar fashion
from a finely resolved electronic density of states, calculated for different volumina with
a Monkhorst-Pack k-grid of $20 \times 20 \times 20$ and Brillouin zone intergration via
the tetrahedron method with Bl\"ochl corrections~\cite{bloechl:94}.
The finite temperature modelling of the electronic subsystem simply involved the folding of the
density of states with the Fermi distribution function. The impact of finite temperature
magnetic spin-flip or spin-wave excitations on the electronic structure,
which was incorporated in the approach of
D\'eak et al.~\cite{deak:14} and Polesya et al.~\cite{polesya:16},
has not been taken into account here. 

\subsection{Samples and experimental procedures}
\label{sub:exp}
Two FeRh thin-film samples (labeled FeRh02 and FeRh03, respectively)
with different stoichiometries were grown by molecular-beam epitaxy
(MBE) via codeposition of $^{57}$Fe-metal and Rh in ultrahigh vacuum onto
clean MgO(001) substrates held at 300$\,^{\circ}$C during deposition. The
preselected deposition rates for $^{57}$Fe (enriched to 95\,\% in the isotope
$^{57}$Fe) and Rh were measured and controlled by several independent
quartz-crystal oscillators. The FeRh film
thickness was about 100 nm. After deposition, the films were in-situ
annealed at 800$\,^{\circ}$C (sample FeRh02) or 700$\,^{\circ}$C (sample FeRh03) in order
to promote the B2 order. The B2 structure and the epitaxial (001)
growth were verified by ex-situ conventional $\Theta - 2 \Theta$ X-ray
diffraction. The actual composition of the samples was inferred from
energy-dispersive X-ray spectroscopy (EDX) and X-ray photoelectron spectroscopy (XPS). The composition was
found to be ($51.4 \pm 1.2$) at.\% Fe for sample FeRh02 (i.e. Fe$_{51}$Rh$_{49}$) and ($48.0 \pm 1.0$) at.\% Fe for sample FeRh03 (i.e. Fe$_{48}$Rh$_{52}$), as compared to
the nominal composition (according to the quartz-crystal oscillators)
of 51 at.\% Fe and 48 at.\% Fe, respectively. Structural details were
studied by high resolution X-ray diffraction using a 4-circle
diffractometer. This diffractometer allowed the determination of the
out-of-plane ($c$) and in-plane ($a$) lattice parameters of the FeRh thin
films by measuring asymmetric reflections, thus having an in- and out-of-plane component. We obtained a $\nicefrac{c}{a}$ ratio of 1.0114(7) for sample FeRh02 (Fe$_{51}$Rh$_{49}$)
and 1.0057(7) for sample FeRh03 (Fe$_{48}$Rh$_{52}$), showing that the $\nicefrac{c}{a}$ ratio is
slightly larger for sample Fe$_{51}$Rh$_{49}$ than for sample Fe$_{48}$Rh$_{52}$.
Our room temperature (RT) $\nicefrac{c}{a}$ values are in good agreement with experimental $\nicefrac{c}{a}$ values of 1.016 for
FM Fe$_{49}$Rh$_{51}$ and 1.008 for AF Fe$_{49}$Rh$_{51}$ epitaxial thin films on MgO(001)
reported by Bordel et al.~\cite{bordel:12}.
The samples were further characterized by ex-situ $^{57}$Fe conversion-electron
M\"ossbauer spectroscopy (CEMS) and vibrating sample magnetometry (VSM). These
results showed that sample FeRh02 (Fe$_{51}$Rh$_{49}$) is ferromagnetic (FM)
from RT down to 5\,K, while sample FeRh03 (Fe$_{48}$Rh$_{52}$)
is antiferromagnetic (AF) up to \unit[$\sim360$]{K}, where it starts to transform
upon heating to the FM state, the transition being completed at \unit[$\sim404$]{K}.
Further details on sample preparation and characterization will be
published elsewhere~\cite{salamon:unpub}.
Shortly (\unit[$<1$]{month}) after sample preparation, the $^{57}$Fe NRIXS measurements were performed
at \unit[$\sim60$]{K}, \unit[$\sim300$]{K} and \unit[$\sim416$]{K} at the undulator beamline 3-ID at the Advanced Photon Source, Argonne National
Laboratory. $^{57}$Fe NRIXS is selective to the $^{57}$Fe resonant
isotope only and measures the phonon excitation probability, as described
in Refs.~\onlinecite{seto:95,sturhahn:95,chumakov:95,chumakov:99}.
This provides the Fe-projected (partial) phonon (or
vibrational) density of states (VDOS) rather directly with a minimum of
modeling~\cite{sturhahn:00}. A high resolution monochromator~\cite{toellner:00} was used to produce x-ray with 
meV energy bandwidth for phonon studies. The monochromatized synchrotron radiation was incident
onto the thin-film surface under a grazing angle of a few degrees. The
synchrotron beam energy was scanned around the resonant energy of the
$^{57}$Fe nucleus (14.413 keV) with an energy resolution $\Delta E$ of 1.3 meV and
was focused onto the sample surface by a Kirkpatrick-Baez mirror. The
average collection time per NRIXS spectrum was about \unit[6--8]{h}. The
evaluation of the NRIXS spectra and the extraction of the VDOS was
performed using the PHOENIX software by W. Sturhahn~\cite{sturhahn:00}.

\section{Harmonic lattice vibrations and structural stability}
\label{sec:structure}

\subsection{Results}
\label{sub:Results}
\begin{figure}[htbp]
	\centering
        \begin{subfigure}[b]{0.4\linewidth}
        \centering
                \includegraphics[width=\textwidth]{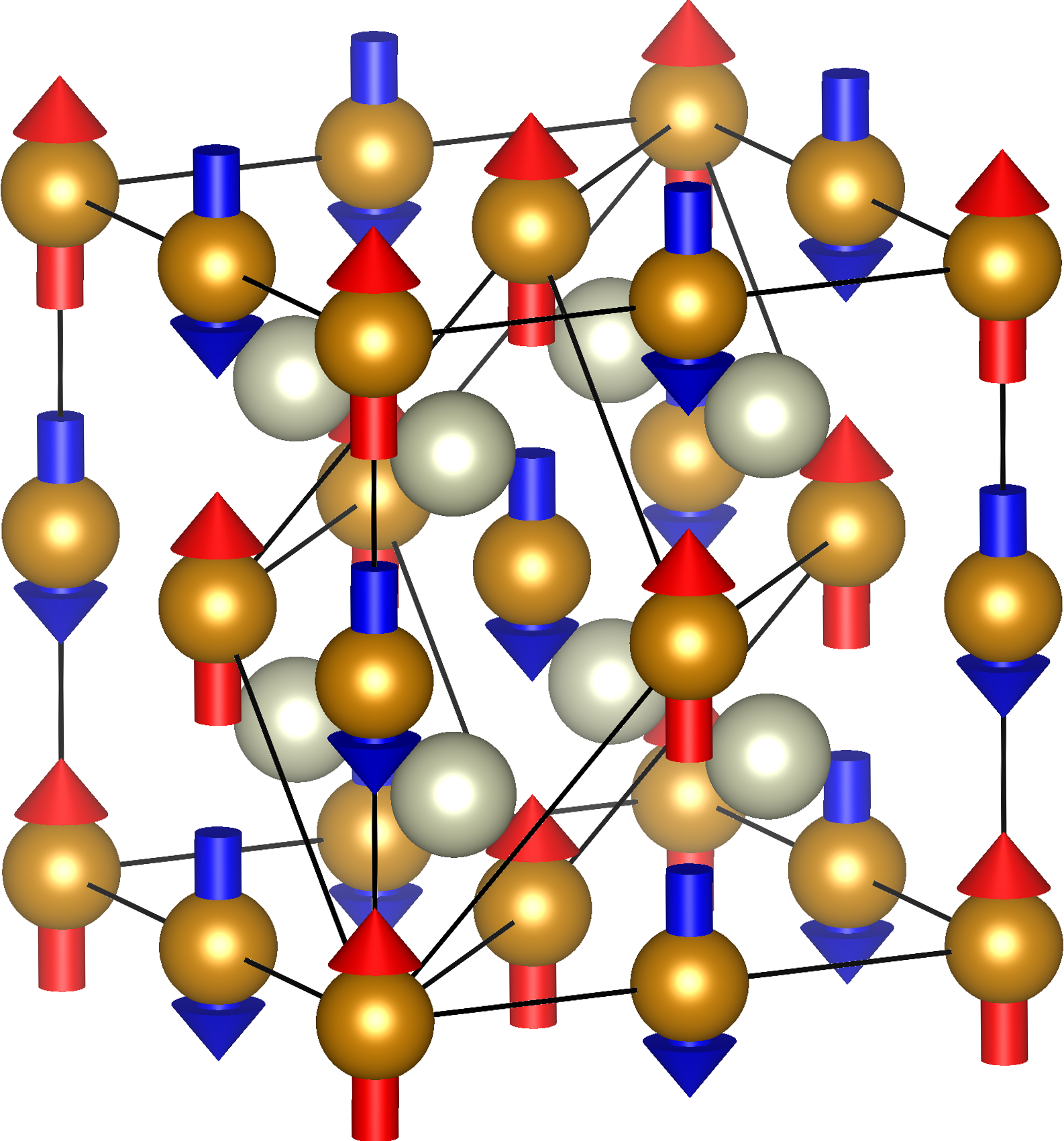}
                \caption{}
                \label{fig:Cells_a}
        \end{subfigure}        
        \begin{subfigure}[b]{0.4\linewidth}
        \centering
                \includegraphics[width=\textwidth]{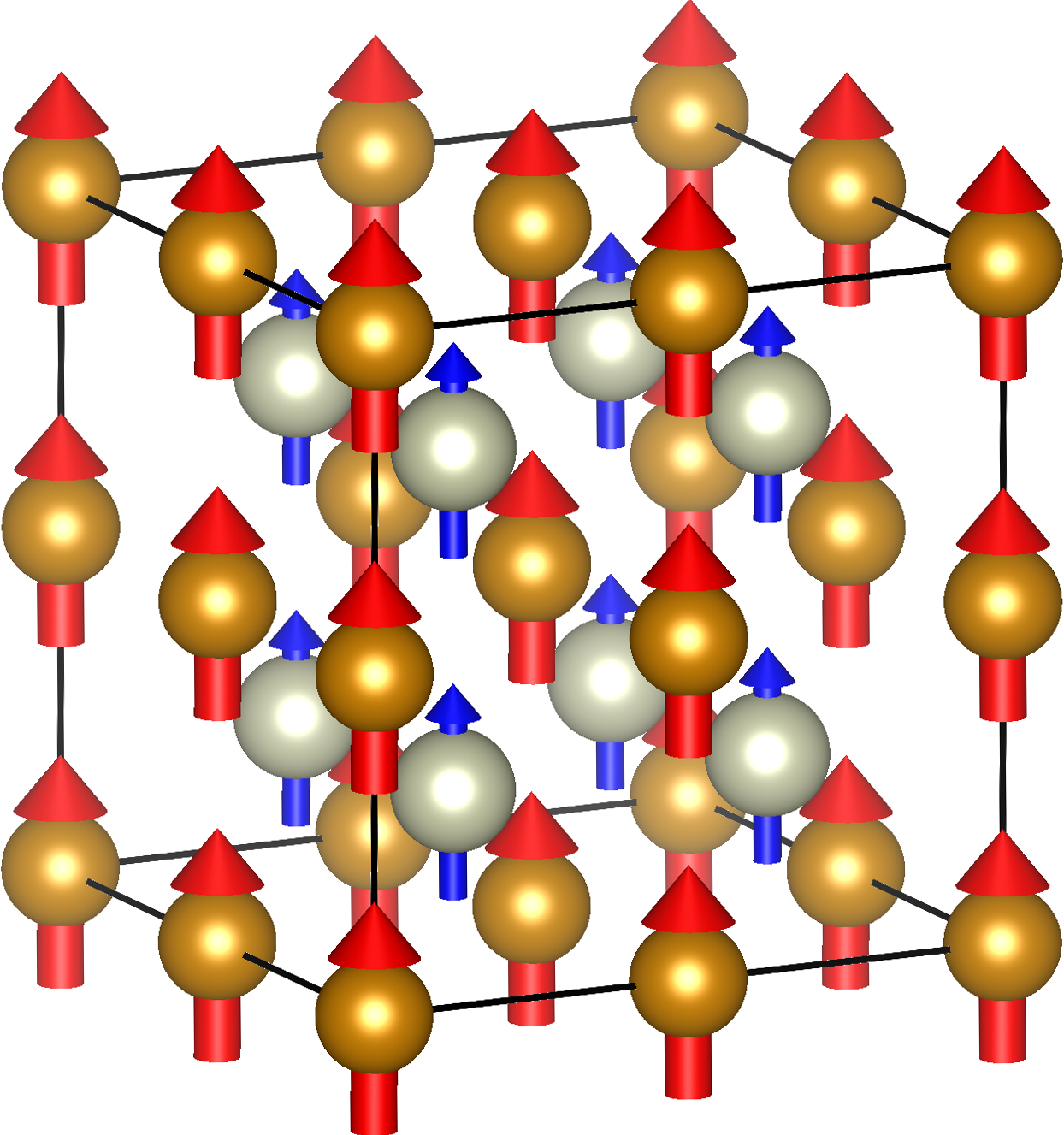}
                \caption{}
                \label{fig:Cells_b}
        \end{subfigure}
    \caption{Magnetic configurations of FeRh. Iron atoms are shown in gold, rhodium in silver. (a) B2(AF) (also called \mbox{AFM-II} or \mbox{AFM-G}) configuration with the fcc-like unit cell shown. Fe atoms in the {111} lattice planes are alternating spin-up (red) and spin-down (blue). (b) B2(FM) state where also the Rh atoms possess a magnetic moment (blue, smaller than the iron moments which are shown in red).}
	\label{fig:Cells}
\end{figure}

In the ordered cubic B2 structure (CsCl prototype) FeRh exhibits two major magnetic configurations, AF coupling (between the $\lbrace111\rbrace$ planes) and FM coupling, which are close in energy and are visualized in Fig.~\ref{fig:Cells}.
Initially calculations where carried out with the PBE GGA functional, and, in accordance with literature, the magnetic ground state for FeRh was found to be the AF configuration, where the Fe atoms are coupled ferromagnetically within the {111} planes with alternating alignment of the spins in adjacent planes. In this configuration the rhodium atoms do not carry a magnetic moment. The FM configuration is slightly higher in energy (\unit[+70.8]{meV/f.u.} compared to AF) and has a larger equilibrium volume. Here, the rhodium atoms also possess a magnetic moment, which appears to be induced by the Fe atoms. Details can be found in table~\ref{tab:Energies}.

\begin{table}[hbt]
\caption{Energy difference to the B2(AF) structure, lattice constants, cell volume, and local magnetic moments of the two studied magnetic configurations of FeRh. Energies and volumes are given per formula unit.}
\label{tab:Energies}
\begin{ruledtabular}
\begin{tabular}{lccccc}
 & $\Delta E$ [meV] & $a$ [\AA] & $V$ [\AA$^3$] & $m^\mathrm{loc}_\mathrm{Fe}$ [$\mu_\mathrm{B}$]& $m^\mathrm{loc}_\mathrm{Rh}$ [$\mu_\mathrm{B}$] \\
 \noalign{\vskip 1mm}
\hline
\noalign{\vskip 1mm}
B2(AF)	  & 0    & 2.990 & 26.73 &  $\pm3.118$ & 0    \\
B2(FM)    & 70.8 & 3.007 & 27.20 &    3.177    & 1.058 \\
\end{tabular}
\end{ruledtabular}
\end{table}

Our ground state lattice constant of $a_\mathrm{AF}=\unit[2.990]{\AA}$ at \unit[0]{K} is in good agreement with the experimental values (\unit[2.986]{\AA}~\cite{shirane:64}, \unit[2.993]{\AA}~\cite{makhlouf:94}, and \unit[3.000]{\AA}~\cite{ibarra:94}) and previous calculations (\unit[2.996]{\AA}~\cite{gu:05}, \unit[2.998]{\AA}~\cite{jekal:15}, and \unit[3.002]{\AA}~\cite{gruner:03,aschauer:16}). The calculated lattice parameter for the FM structure at \unit[0]{K} is, at $a_\mathrm{FM}=\unit[3.007]{\AA}$, also in good agreement with previous work (\unit[3.020]{\AA}~\cite{gruner:03}, \unit[3.012]{\AA}~\cite{jekal:15}, and \unit[3.018]{\AA}~\cite{gu:05,aschauer:16})

Earlier investigations reported that the B2(AF) phase is soft with
respect to a tetragonal distortion corresponding to the martensitic
Bain path from the body centered cubic (bcc) to the face centered
cubic (fcc) structure~\cite{uebayashi:06,cherifi:14,aschauer:16}.
Investigating the Bain path and
optimizing the volume of the cell at each step we confirm a second
minimum at $\nicefrac{c}{a}=1.247$ (pure fcc:
$\nicefrac{c}{a}=\sqrt{2}\simeq1.414$) after overcoming a barrier of
only \unit[$\sim2$]{meV/f.u.}\ at $\nicefrac{c}{a}\simeq1.1$.\footnote{In
contrast to Ref.~\onlinecite{aschauer:16} we find this distorted
structure to be slightly
lower in energy by \unit[$\sim 5$]{meV/f.u.} than the
bcc phase using the PBE functional. This result did not change after
increasing the plane wave cutoff to \unit[550]{eV} as used in
Ref.~\onlinecite{aschauer:16}. For the pure fcc phase the energy is
higher than bcc by \unit[74.4]{meV/f.u.}\ at zero pressure.}
Indeed, X-ray diffraction experiments found a transition to a mixture of body
centered tetragonal (bct) and fcc under a small uniaxial pressure of
\unit[0.25]{GPa} and by
admixture of Pt or Pd,\cite{miyajima:92,yuasa:94,yuasa:95,uebayashi:07}
and the careful analysis of the elastic behavior in
Ref.\ \onlinecite{aschauer:16} reveals that the tetragonal
shear constant $C^{\prime}$ exhibits a marked softening under compressive
volumetric and epitaxial strain.
A structural transition was also observed in thin films of
disordered FeRh by Witte and coworkers.\cite{witte:16}

\subsubsection{Phonon calculations}
\label{subsub:phonons}

\begin{figure}[htbp]
	\centering
    \includegraphics[width=\linewidth]{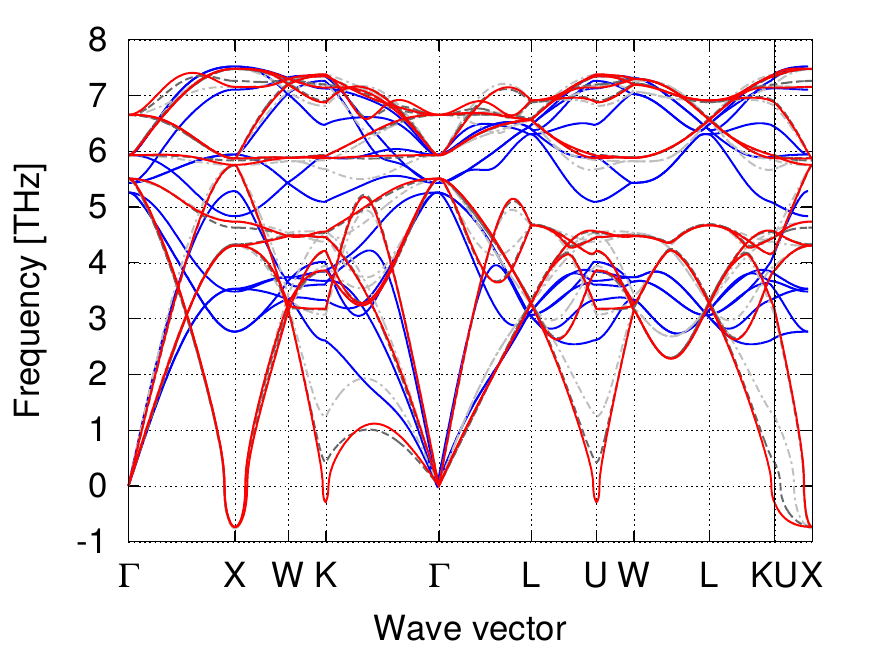}
    \caption{Phonon band structures for B2 FeRh. Solid red (blue) lines are for $4\times4\times4$ supercells in the AF (FM) magnetic configuration. Light grey dashed-dotted lines are results from $2\times2\times2$, dark grey dashed lines are for $3\times3\times3$ AF supercells. Imaginary frequencies are plotted as negative.}
	\label{fig:Phonons}
\end{figure}

To determine the phononic contribution to the magnetic phase transition described in section~\ref{sec:Intro} we calculated the phonon band structure for the two cubic magnetic phases, FM and AF. For better comparison we used the same primitive cell with fcc basis vectors sketched in Fig.~\ref{fig:Cells_a} for both magnetic phases. Convergence of the stable branches with respect to the supercell size was achieved at $4\times4\times4$ multiplication of the 4 atom unit cell, which then contains 256 atoms (see Fig.~\ref{fig:Phonons}).

While the B2(FM) dispersion is stable in the entire Brillouin zone, a region around the $X$ point of the B2(AF) phonon band structure shows imaginary (plotted as negative) frequencies for all cell sizes (see Fig.~\ref{fig:Phonons}). This indicates a dynamic instability of the crystal and suggests that displacing the ions according to the wave vector at $X$ would not result in a restoring force but lead to a lowering of the total energy. The wave vector at $X$, which points along the direction of one of the cubic axes, describes a (doubly degenerate) transverse optical phonon with a periodicity of $2a_\mathrm{AF}$. Comparing results for different supercell sizes shows that the instability does not depend on the cell size along the $\Gamma - X - W$ direction of the Brillouin zone, but spreads out along $X - U$ and $X - K$ with increasing cell size until also $U$ and $X$ are included in the imaginary pocket for the $4\times4\times4$ supercell.

Phonon calculations require very accurate forces and thus we repeated our volume optimization and phonon calculations with significantly higher plane wave cutoff (\unit[850]{eV}) and k-mesh density (corresponding  to a $24\times24\times24$ mesh for the unit cell). The phonon band structure obtained with these parameters showed no significant deviation from Fig.~\ref{fig:Phonons}, ruling out insufficiently converged computational parameters as a source of the imaginary frequencies.

\begin{figure}[htbp]
	\centering
        \begin{subfigure}[b]{0.45\linewidth}
                \includegraphics[width=\textwidth]{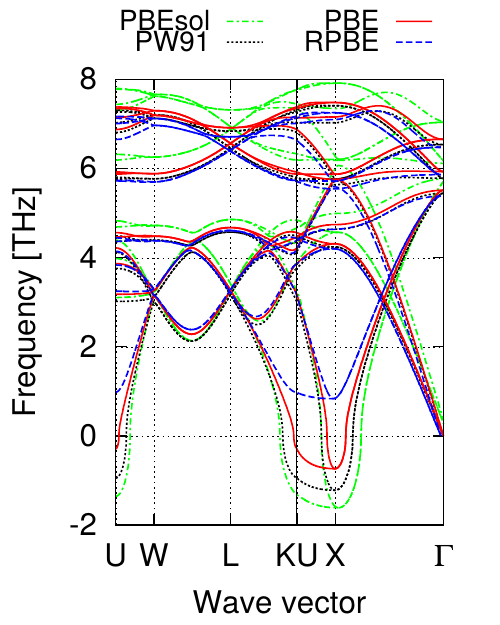}
                \caption{different GGAs}
                \label{fig:Phonons_GGA}
        \end{subfigure}                
        \begin{subfigure}[b]{0.45\linewidth}
                \includegraphics[width=\textwidth]{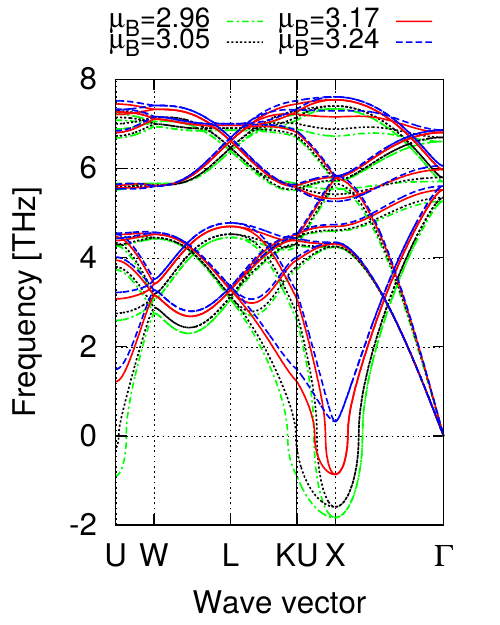}
                \caption{constrained PBE}
                \label{fig:Phonons_Constrained}
        \end{subfigure}
    \caption{(a) Phonon band structures for B2(AF) FeRh calculated for different GGAs. All GGAs other than RPBE show dynamical instabilities. (b) Qualitative PBE phonon band structures for B2(AF) FeRh calculated for constrained magnetic moments. Increasing the magnetic moments on the Fe atoms from \unit[$\pm 2.96$]{$\mu_\mathrm{B}$} (green dash-dotted line) to \unit[$\pm 3.24$]{$\mu_\mathrm{B}$} (blue dashed line) gradually stabilizes the whole band structure. If dotted, dash-dotted or dashed lines appear solid, it is due to degenerate bands.}
     \label{fig:Phonons_Constrained+GGA}
\end{figure}

 To investigate whether the instability depends on the applied functional (PBE) we performed additional calculations with other GGAs, first computing the equilibrium volume and then determining the phonon band structure for PBEsol, PW91, and RPBE. The relevant parts of these phonon band structures are plotted in Fig.~\ref{fig:Phonons_GGA} and significant differences are revealed for the four tested GGA functionals.

Although the stable phonon branches (both acoustic and optical) are in good qualitative agreement, the phonon instability previously detected for PBE is not a universal feature. For PBEsol and PW91, the imaginary pocket around the $X$ point is significantly broader and deeper clearly encompassing also the $U$ and $K$ points, where the PBE bands remain nearly stable. On the other hand, the phonon band structure calculated with the RPBE functional is stable at all wave vectors although it also shows considerable softening of the transverse optical modes at the $X$, $K$, and $U$ points. To further analyze this unsatisfactory result we performed additional volume optimization and phonon calculations for the local density approximation (LDA) and the van der Waals corrected optB86b functionals, as well as a PBE phonon calculation at the higher volume predicted by the RPBE functional. Both the LDA and the optB86b calculations lead to a smaller volume and smaller magnetic moments than all GGAs and both lead to large imaginary pockets around $X$ encompassing also $U$ and $K$, while the PBE calculation at the higher volume (PBE* in Tab.~\ref{tab:Functionals}) leads to an increased magnetic moment and has only a very small pocket at the $X$ point in the phonon band structure.
The results of all calculations are given in table~\ref{tab:Functionals}, sorted from top to bottom according to increasing magnetic moments. We see (also note Fig.~\ref{fig:Phonons_GGA}) that an increase in magnetic moment is stabilizing the B2(AF) phase, as the imaginary pocket is reduced in size from PBEsol (\unit[$\pm3.04$]{$\mu_\mathrm{B}$}) and PW91 (\unit[$\pm3.10$]{$\mu_\mathrm{B}$}), over PBE (\unit[$\pm3.12$]{$\mu_\mathrm{B}$}), to PBE* (\unit[$\pm3.15$]{$\mu_\mathrm{B}$}), and finally vanishes for RPBE (\unit[$\pm3.19$]{$\mu_\mathrm{B}$}).

Comparing the GGA results, the increased magnetic moment is correlated with an increased atomic volume.
For instance,  RPBE, at \unit[27.4]{\AA$^3$/f.u.}, predicts a significantly higher volume than PBE (\unit[26.7]{\AA$^3$/f.u.}) and PW91 (\unit[26.8]{\AA$^3$/f.u.}). In turn, a calculation with PBE at an artificially higher volume
(labeled PBE* in Tab.~\ref{tab:Functionals}) also led to a nearly stable phonon band structure. 
To separate the influence of the magnetic moment and volume on the imaginary mode, we performed PBE
phonon calculations with constrained magnetic moments using $2\times2\times2$ supercells at the equilibrium
PBE volume of \unit[26.7]{\AA$^3$}.
Although we were not able to converge the constrained calculations to the same accuracy as our other phonon calculations,
Fig.~\ref{fig:Phonons_Constrained} indicates that the increase of the magnetic moments alone is sufficient to stabilize the cubic structure. This observation is in accordance with the results by Aschauer et al.~\cite{aschauer:16},
who find that the second minimum along the martensitic Bain path disappears if the local Fe moments are increased.
In contrast to our work, however, they use a PBE+U approach to increase electron localization in the Fe $d$ orbitals and thus influence the magnetic moments, while we subtly change the hybridization of the orbitals by increasing the size of the magnetic moments directly.

\begin{table}[htbp]
\caption{\label{tab:Functionals} Comparison of lattice parameter $a$, cell volume $V$, local magnetic iron moments $m^\mathrm{loc}_\mathrm{Fe}$, and phonon stability for the calculation of FeRh in the cubic AF phase with different functionals. PBE* is a PBE phonon calculation at higher (non-equilibrium) volume.}
\begin{ruledtabular}
\begin{tabular}{lcccccc}
 & $a$ [\AA] & $V$ [\AA$^3$/f.u.] & $m^\mathrm{loc}_\mathrm{Fe}$ [$\mu_\mathrm{B}$]& Instabilities at\\
\noalign{\vskip 1mm}
\hline
\noalign{\vskip 1mm}
LDA & 2.915 & 24.77 & $\pm2.835$ &  $X$, $U$, $K$ \\
optB86b & 2.965 & 26.07 & $\pm3.036$ & $X$, $U$, $K$ \\
PBEsol & 2.947 & 25.60 & $\pm3.040$ &  $X$, $U$, $K$ \\
PW91 & 2.992 & 26.78 & $\pm3.095$ & $X$, $U$, $K$ \\
PBE & 2.990 & 26.73 & $\pm3.117$ & $X$, $U$, $K$ \\
PBE* & 3.014 & 27.39 & $\pm3.152$ &  $X$ \\
RPBE & 3.014 & 27.39 & $\pm3.183$ &   \\
\noalign{\vskip 1mm}
\hline
\noalign{\vskip 1mm}
Exp.~\cite{shirane:64} & 2.986 & 26.63 & $\pm3.3$ & - \\
Exp.~\cite{makhlouf:94} & 2.993 & 26.81 & - & - \\
Exp.~\cite{ibarra:94} & 3.000 & 27.00 & - & - \\
\end{tabular}
\end{ruledtabular}
\end{table}

While it is interesting that quite small differences in the magnetic moments can lead to a stabilization of the cubic phase, it is certainly true that the size of the magnetic moments and the volume are interrelated effects and an increase of one of them leads to an increase in the other unless some degree of freedom is constrained. Indeed, if the PBE volume is increased to \unit[27.39]{A$^3$} the magnetic moments converge to \unit[$\pm 3.15$]{$\mu_\mathrm{B}$} (PBE* in Tab.~\ref{tab:Functionals}) and the unstable pocket at $X$ is considerably smaller than the one observed if the volume is not increased and only the magnetic moment is constrained to \unit[$\pm 3.17$]{$\mu_\mathrm{B}$} (see figure~\ref{fig:Phonons_Constrained}).

\subsubsection{Analyzing the phonon instability}
\label{subsub:instability}

\begin{figure}[htbp]
	\centering
    \includegraphics[width=0.85\linewidth]{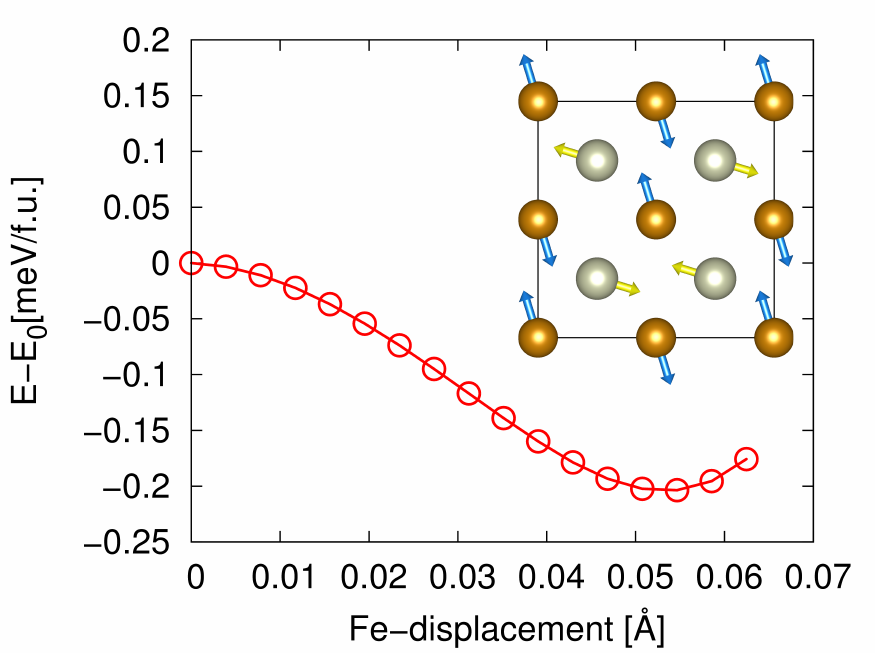}
    \caption{Frozen phonon calculation corresponding to the unstable phonon at wave vector $X$. Energy is referenced to zero displacements. The displacement is given only for the Fe atoms since it is not equivalent for Fe and Rh. The inset shows the cubic unit cell and direction of displacements of ions according to the dynamic instability at the $X$ point. Arrows indicate the direction of the displacement, Fe is shown in gold, Rh in silver.}
     \label{fig:frozen_phonon}
\end{figure}

We already mentioned that the instability at the $X$ point in the PBE phonon band structure (see Fig.~\ref{fig:Phonons}) is a doubly degenerate transverse optical phonon branch with the wave vector pointing along one of the cubic axes. Without loss of generality we chose the $c$ axis as direction of the wave vector. The direction of the ionic displacements resulting from this wave vector is shown in the inset of Fig.~\ref{fig:frozen_phonon}. The angle of the ion displacements with the cubic axes of the cell is $\sim 17^\circ$ and the Rh atoms become displaced about 16\% less than the Fe atoms. 
By stepwise displacing the atoms without allowing cell or atomic relaxations, we see a slight lowering of the total energy and a minimum at an amplitude of about \unit[0.05]{\AA} (see Fig.~\ref{fig:frozen_phonon}).

\begin{figure}[htbp]
	\centering
		\begin{subfigure}[b]{0.49\linewidth}
			\centering
 			\includegraphics[width=\textwidth]{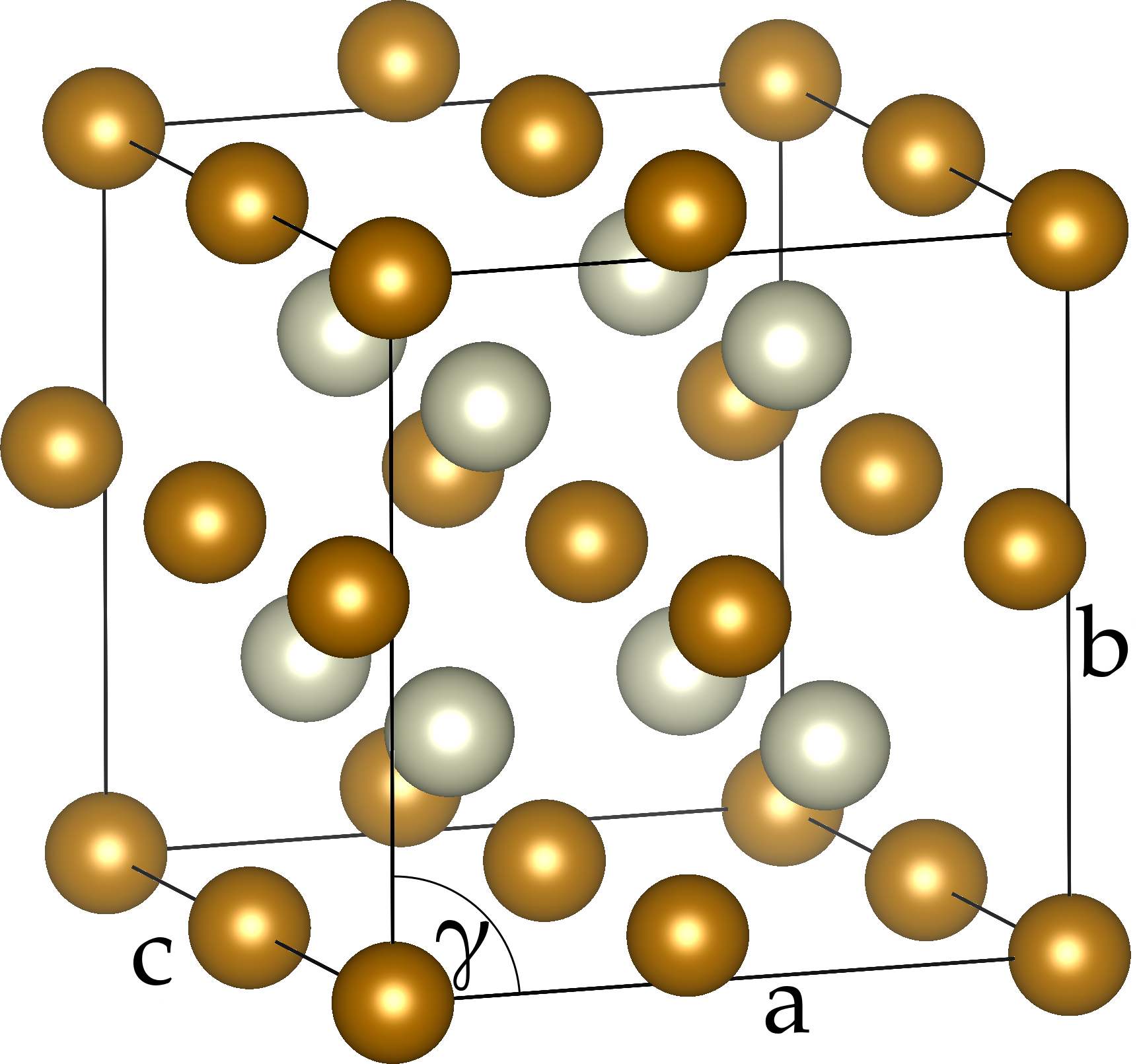}
        	\caption{}
            \label{fig:monoclinic_cell_3d}
        \end{subfigure}        
        \begin{subfigure}[b]{0.49\linewidth}
	        \centering
            \includegraphics[width=\textwidth]{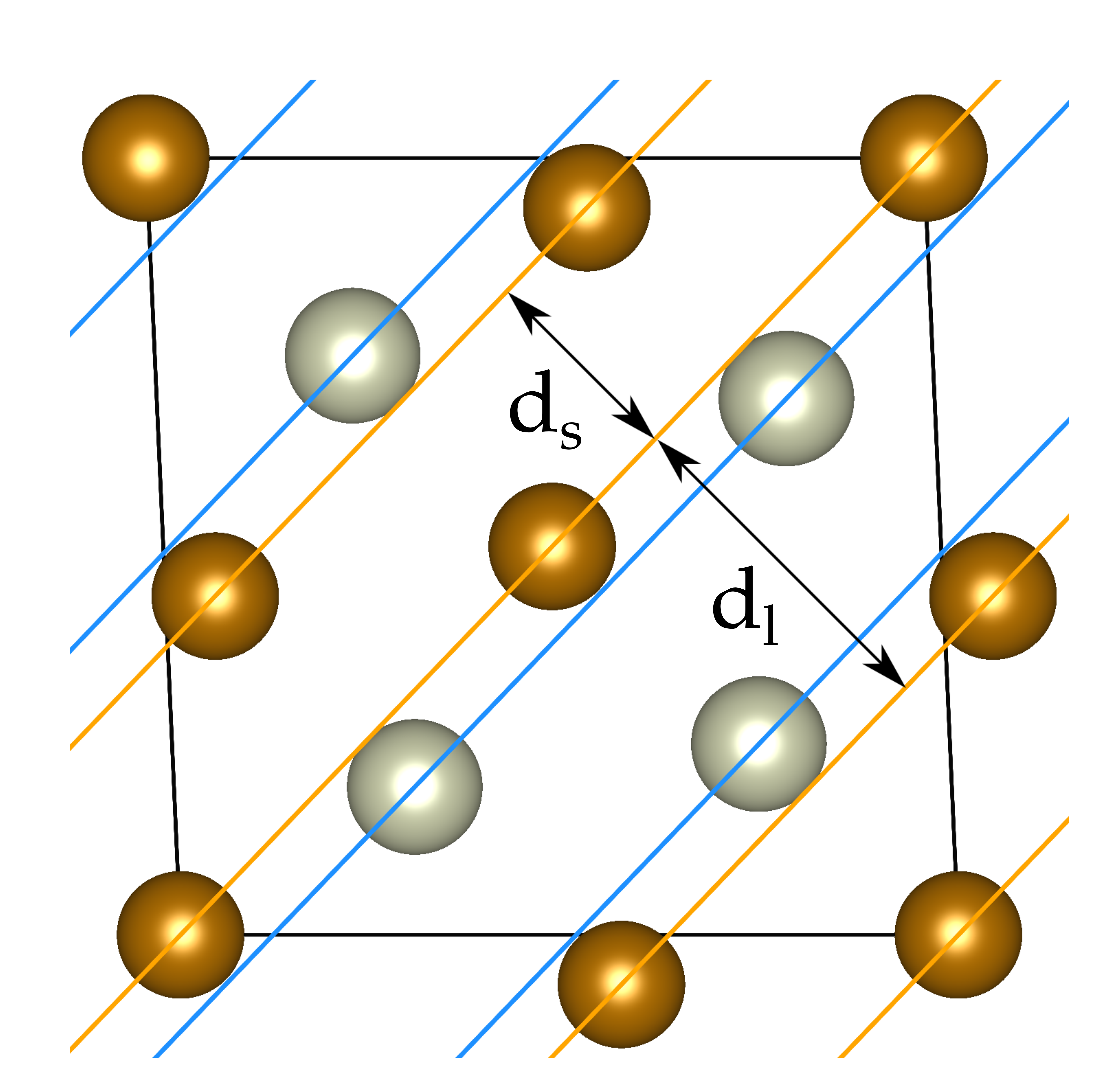}
            \caption{}
            \label{fig:monoclinic_cell_planes}
        \end{subfigure}
        \caption{(a) Monoclincic unit cell of $P$2/$m$(AF) FeRh. (b) Side view of the same cell with $\lbrace 1  \overline{1} 0\rbrace$ planes drawn for both Fe (gold) and Rh (silver) atoms. Arrows indicate short ($d_s$) and long ($d_l$) distance between lattice planes.}
     \label{fig:monoclinic_cell}
\end{figure}

While the total reduction in energy is only \unit[$\sim0.2$]{meV} per formular unit, this minimum confirms the phonon calculations and can be used as a starting point for cell- and subsequent ionic relaxations.
After carefully relaxing the whole system we arrive at a monoclinic structure ($P$2/$m$) with a total energy gain of \unit[24.3]{meV} per formular unit compared to B2(AF) (see Fig.~\ref{fig:monoclinic_cell})\footnote{It is to be noted that the energy landscape of FeRh is rather flat with respect to these atomic displacements and lattice distortions, thus finding the minimum for the monoclinc phase proved to be a rather difficult endeavor, requiring a lot of different relaxation steps and combination of algorithms.}.
This is more than twice the energy gain reported for a tetragonally distorted structure with similar atomic displacements in Ref.~\onlinecite{aschauer:16}.

The change in the angle $\gamma$ to $92.6^\circ$ is not particularly large, but the tetragonal distortion is severe, with $c$ compressed by 7.2\% to \unit[5.55]{\AA} and $a$ and $b$ enlarged by 2.8\% to \unit[6.15]{\AA}. The distortion of the lattice vectors leads to a reduction in volume by $\sim 2\%$. The magnetic moments on the Fe sites are also reduced by $\sim 9\%$ to \unit[$\pm2.816$]{$\mu_\mathrm{B}$} for the $P$2/$m$(AF) phase compared to the B2(AF) structure. The Rh atoms still do not carry a local moment, a result that was carefully checked by turning all symmetry operations off. In Fig.~\ref{fig:monoclinic_cell_planes} the $\lbrace 1 \overline{1} 0\rbrace$ lattice planes are drawn to clarify in which way the ions are shifted compared to the cubic structure. If one compares Fig.~\ref{fig:monoclinic_cell_planes} with the cubic structure (inset of Fig.~\ref{fig:frozen_phonon}), it becomes clear that the $\lbrace 1 \overline{1} 0\rbrace$ lattice planes, which contain both Fe and Rh atoms in the cubic phase, contain only a single atomic species in the monoclinic phase. Now two closer spaced planes of Fe atoms are followed by two closer spaced planes of Rh atoms, and so forth. The distances between Fe planes (short: $d_s=\unit[1.65]{\AA}$; long: $d_l=\unit[2.79]{\AA}$; for Rh planes the situation is exactly the same), indicated by arrows in Fig.\ref{fig:monoclinic_cell_planes}, show the deviation from the cubic structure, where those lattice planes are equidistant with $d_0=\unit[2.11]{\AA}$.

\begin{figure}[htbp]
	\centering
    \includegraphics[width=1\linewidth]{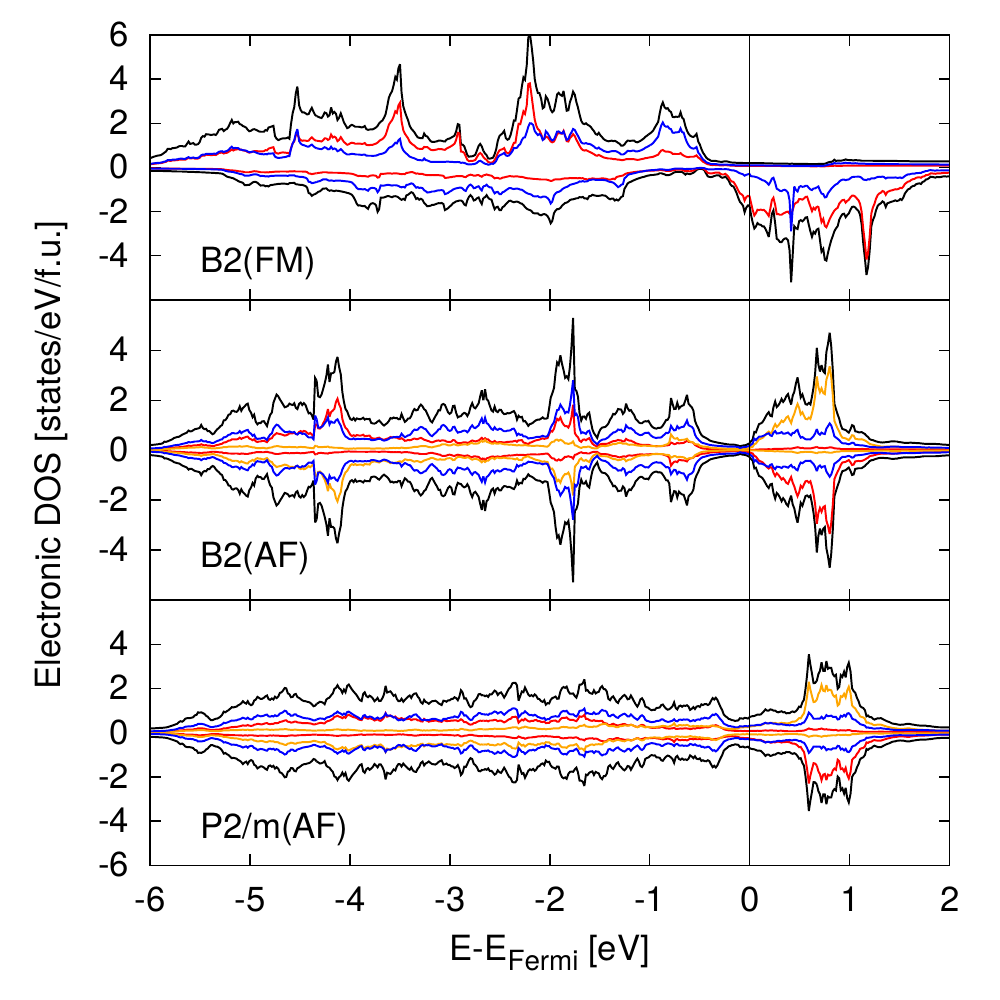}
    \caption{Electronic DOS for the B2(FM), B2(AF) and  $P$2/$m$(AF) phases of FeRh. Total DOS is plotted in black, Fe states are red and orange (depending on the nonequivalent Fe sites in the AF structure) and Rh states are blue.}
     \label{fig:DOS}
\end{figure}

Considering the density of states (DOS) for both crystal structures (see Fig.~\ref{fig:DOS}), we observe a significant increase of the AF DOS at the Fermi level from \unit[0.6]{states/eV/f.u.} for the B2 case to \unit[1.4]{states/eV/f.u.} for the $P$2/$m$ phase, although it is still smaller than the DOS at the Fermi level in the B2(FM) case (\unit[1.8]{states/eV/f.u.}). The loss of symmetry is also clearly visible in the monoclinic DOS, which loses most of the distinct features displayed in the cubic phase and is essentially uniform in the valence band.

We also investigated the stability of the monoclinic phase with respect to other functionals. To this end we optimized the volume of the monoclinic cell for all functionals in table~\ref{tab:Functionals} which show the dynamic instability, while fixing the cell shape and holding the ion positions at their PBE relaxed coordinates. The monoclinic phase is significantly favored in energy for all the functionals, with LDA showing the strongest decrease in energy and PBE the weakest (see Tab.~\ref{tab:monoclinic}). In all cases the volume and the magnetic moments are significantly reduced, with an average reduction of the volume by 2.2\% and of the local iron moments by 10.9\% for the 3 GGAs PBEsol, PBE, and PW91.

\begin{table}[htbp]
\caption{\label{tab:monoclinic} Energy gain $\Delta E$, volume change $\Delta V$, and Fe local magnetic moment change $\Delta m^\mathrm{loc}_\mathrm{Fe}$ of the $P$2/$m$(AF) structure compared to the B2(AF) phase for different functionals. Cell shape and ion positions have been relaxed with PBE, but volumes are optimized for each functional.}
\begin{ruledtabular}
\begin{tabular}{lccc}
 & $\Delta E$ [meV/f.u.] & $\Delta V$ [\%] & $\Delta m^\mathrm{loc}_\mathrm{Fe}$ [\%] \\
\noalign{\vskip 1mm}
\hline
\noalign{\vskip 1mm}
LDA & $-193.1$ & $-3.4$ & $-22.8$ \\
PBEsol& $-87.6$ & $-2.5$ & $-13.28$ \\
optB86b & $-82.3$ & $-5.0$ & $-13.6$ \\
PW91 & $-42.0$ & $-2.1$ & $-10.1$ \\
PBE & $-24.3$ & $-1.9$ & $-9.2$ \\
\end{tabular}
\end{ruledtabular}
\end{table}

For RPBE, where the B2(AF) structure is stabilized because of the higher magnetic moments and the larger equilibrium volume, the $P$2/$m$(AF) phase is higher in energy by \unit[4.9]{meV/f.u.} if only the volume is optimized according to the RPBE functional. If we also allow the ions and the cell shape to relax again with RPBE, we find that the monoclinic phase is almost equivalent in energy (favored by \unit[-0.7]{meV/f.u.}) to the B2(AF) phase for this functional.

The comparison of the different exchange-correlation functionals proves that from the
computational point-of-view, the preference for the $P$2/$m$(AF) ground state
is a robust result. Indeed, a closely related relaxation pattern as shown
in Fig.\ \ref{fig:frozen_phonon}
(missing the monoclinic distortion) has been found experimentally
in ternary bct FeRh$_{0.38}$Pd$_{0.62}$.\cite{yuasa:95b}
Experiments for binary B2 FeRh applying hydrostatic pressures of
up to \unit[7]{GPa} did not reveal any indications for a new phase
at room temperature and above in pure FeRh.\cite{wayne:68,vinokurova:76}
Later experimental work~\cite{kuncser:05} 
at higher pressures (10 to \unit[20]{GPa})
suggests a transition to a fct tetragonal structure with significantly reduced
volume, which coexists with the B2 phase.

The fact that
the monoclinic phase has not been found in experiments for pure FeRh
until today, although it should be clearly distinguishable from the B2 phase given
the large tetragonal distortion ($\nicefrac{c}{a}=0.9$), requires some discussion.
An early measurement
by neutron diffraction reports a Fe moment of
\unit[3.30]{$\mu_\mathrm{B}$}~\cite{shirane:64},
which is clearly underestimated by all GGA-type exchange-correlation functionals.
According to our constrained moment calculations (see Fig.~\ref{fig:Phonons_Constrained}),
enhancing the Fe moment to \unit[3.30]{$\mu_\mathrm{B}$}
leads to a stable B2(AF) dispersion with standard GGA, even if the lattice constant is
fixed at $a_\mathrm{AF}=\unit[2.990]{\AA}$ corresponding to the experimental
value~\cite{shirane:64,makhlouf:94,ibarra:94}. However,
the GGA is usually known to overestimate magnetic
moments~\cite{asada:92,hafner:02,ruban:08,mazin:08,pulikkotil:12}
(rather than to underestimate them) and increased
magnetic moments would also destabilize the second fcc like minimum
along the martensitic Bain path, which has been experimentally
observed for FeRh under high velocity impact deformation or
filing~\cite{lommel:67,miyajima:92,aschauer:16}. Alternatively,
the extremely shallow minimum associated with the unstable phonon
(only \unit[0.2]{meV/f.u.} energy gain in the cubic phase) may be simply
smeared out by kinetic fluctuations at temperatures larger than
\unit[2]{K}. We also know from first principle calculations that a
small number of antisite defects are enough to suppress AF order down
to low temperatures~\cite{staunton:14}.  It is thus possible that
defects and slightly off-stoichiometric compositions also suppress the
monoclinic AF phase.  Nevertheless, careful adaption of the
$\nicefrac{c}{a}$ ratio or application of strain at low temperatures
could still lead to a stabilization of the $P$2/$m$(AF) low energy
phase with its significantly reduced volume and magnetic moments.

\subsection{Electronic origin of the lattice instability}
\label{sec:FermiSurface}
\begin{figure}[tbh]
\begin{center}
\includegraphics[angle=0, width=0.9\columnwidth]{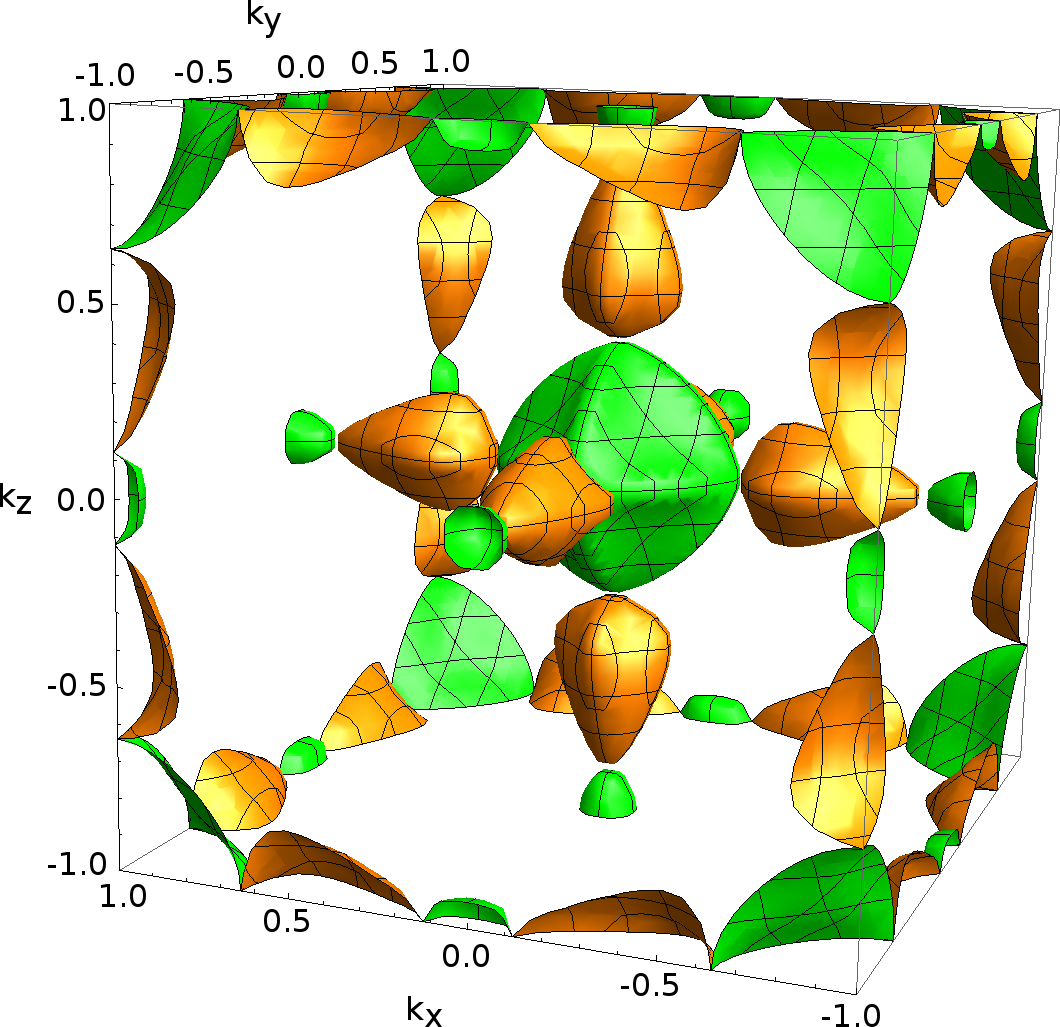}
\end{center}
\caption{Fermi surface of B2-ordered AF FeRh in the extended Brillouin zone scheme.
  The different colors refer
  to different bands crossing the Fermi level. Clear nesting features are absent.
  The shape of the Fermi surface agrees well with previous
  reports.\protect\cite{vinokurova:88,szajek:94}
}\label{fig:FS}
\end{figure}
Taking into account the magnetic configuration, the FeRh lattice in the cubic AF state belongs
to the same point group as the L2$_1$ Heusler alloys
like Ni$_2$MnGa. These compounds similarly exhibit a soft acoustic phonon
in [110] direction, albeit not at the zone boundary point $K$ or even $X$~\cite{uijttewaal:09}.
The origin of the softening in Ni-Mn-based Heusler compounds is a
Ni e$_u$-peak right below the Fermi level, which gives rise
to extended plane nesting sheets in the Fermi surface connected by the same wave vector describing the soft phonon. As pointed out recently, disordered equiatomic bcc-FeRh under epitaxial strain undergoes a transition to an orthorhombic structure which shows signatures of
a martensitic transformation, driven by a redistribution of electronic states away
from the Fermi level in combination with the removal of parallel features visible in selected
Fermi surface cross-sections~\cite{witte:16}.
In B2 FeRh, however, 
the Fermi surface of the AF phase, depicted in Fig.\ \ref{fig:FS}
is rather small, which reflects the low density of states at the Fermi level
(cf.\ Fig.\ \ref{fig:DOS}). It also does not exhibit obvious nesting features.
Lowering the energy of the cubic structure according to the soft phonon mode
following the path displayed in Fig.\ \ref{fig:frozen_phonon} does not lead to notable changes
in the DOS at the Fermi level. Instead, peaks at 1.9\,eV and
4.2\,eV below $E_{\rm_F}$, which arise
from flat hybridized d-bands degenerate at $X$, $L$ and $\Gamma$,
split up asymetrically according to the reduced symmetry. This reminds of a band-Jahn-Teller mechanism
which is taking effect far below the Fermi level.
  The overall gain in energy is very small and it is impossible
  to clearly separate the competing contributions.
  But we may speculate that the strong hybridization of rhodium and iron states
  in both the AF and the FM phase, responsible for both the implicit splitting of Rh
  in the AF phase and the evolution of a net magnetic moment in the
  FM phase,\cite{sandratskii:11,kudrnovsky:15}
  is also decisive for the structural stability of B2 FeRh.
  This could explain why rather different mechanisms such as increasing the local exchange splitting of Fe
  via constrained magnetic moments, or decreasing the band width by increasing the volume, or shifting
  the relative position of the elemental orbitals using the GGA+U approach, have the same
  consequence, i.\,e., stabilizing the B2(AF) phase. 

An instable phonon can open a
downhill relaxation path to a new ground state structure, as the monoclinic structure
for ordered FeRh or the tetragonal L$1_0$ phase in the case of the Heusler
systems~\cite{niemann:12,gruner:14}.
In FeRh, the phonon instability is fragile and is connected to a much smaller energy gain
compared to the Heusler alloys. 
  Thus, in addition to the mechanisms discussed above, the relaxation path to the monoclinic ground state
  might easily be blocked by other perturbations as well, like a slight amount of chemical or
  magnetic disorder, or the presence of lattice defects. However, the
phonon-induced modulations are no prerequisite for the existence of a
(meta-)stable monoclinic minimum. This is directly seen from a comparison
of the electronic density of states (see Fig.~\ref{fig:DOS}), which exhibits entirely
unrelated features for the B2 and P2/m phases, as well as the existence of the minimum for the monoclinic phase for the RPBE functional which predicts stable phonons in B2(AF) structure.
This suggests, that the new phase might be stabilized at low temperatures under
carefully designed external conditions, such as a sufficiently large pressure,
by epitaxial strain and/or band filling, which tunes the
AF-FM transition as well.\cite{barua:13}

\section{Nuclear resonant inelastic X-ray scattering in thin films}
\label{sec:experiment}
\begin{figure}[tbh]
\begin{center}
\includegraphics[angle=0, width=0.85\columnwidth]{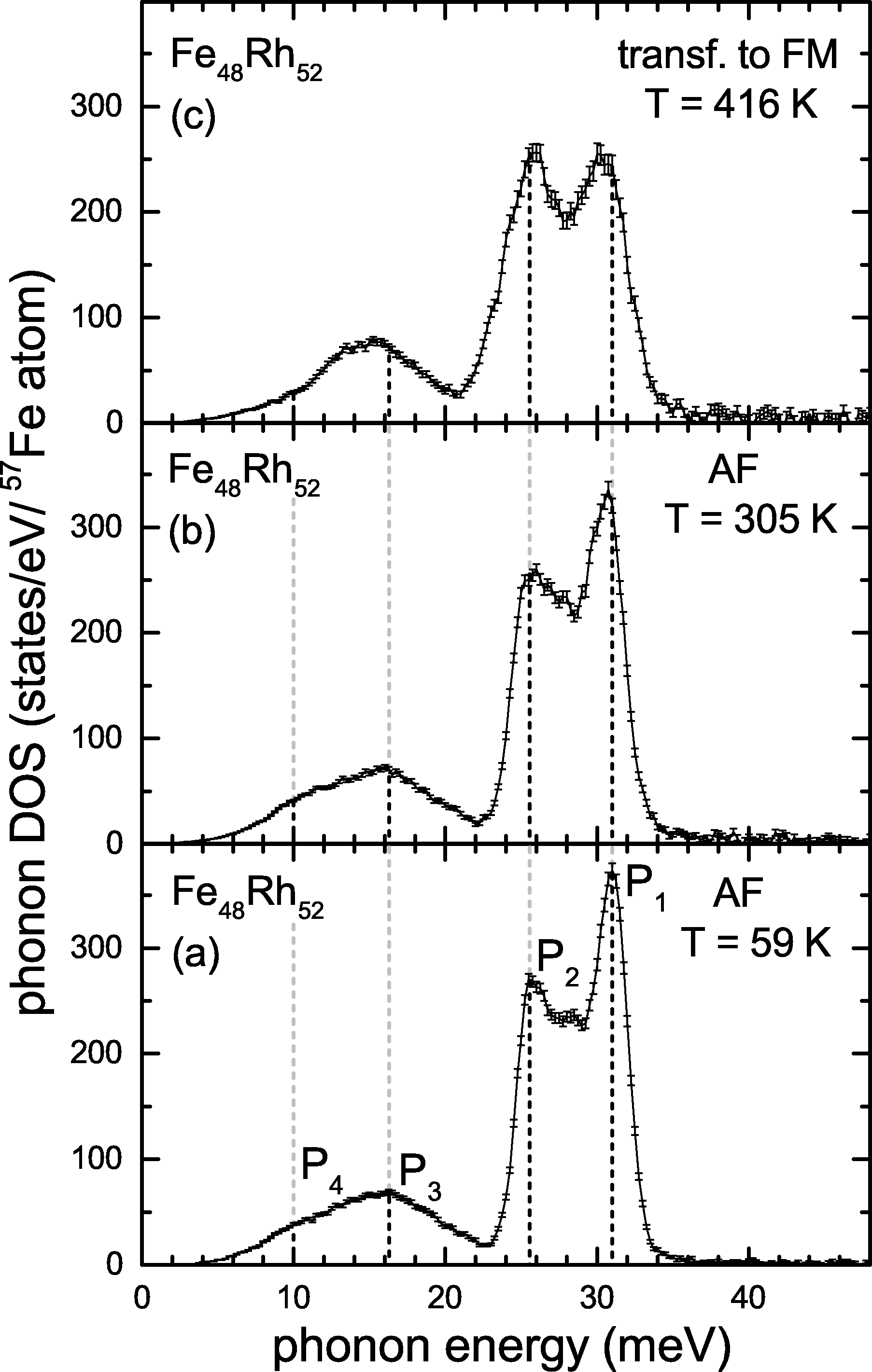}
\end{center}
\caption{
  Fe-projected (partial) VDOS of 
  Fe$_{48}$Rh$_{52}$
  (sample code FeRh03) measured by $^{57}$Fe NRIXS at (a) T = 59\,K (AF), (b) 305\,K (AF),
  and (c) 416\,K (FM). The approximate positions of peaks P$_1$, P$_2$, P$_3$ and of the shoulder
  P$_4$ are indicated by dashed vertical lines. Comparison of the VDOS in (a),
  (b) with (c) reveals distinct differences between the VDOS of the AF and FM state. 
}\label{fig:ExpVDOS}
\end{figure}

Our most important experimental result is presented in Fig.~\ref{fig:ExpVDOS}.
It displays the Fe-projected (partial)
VDOS, $g(E)$, of sample Fe$_{48}$Rh$_{52}$ (code FeRh03) measured at $T = 59$\,K, 305\,K
and 416\,K, i.e., across the AF to FM phase
transition. In Fig.\ \ref{fig:ExpVDOS}(a), the VDOS of Fe$_{48}$Rh$_{52}$
at 59\,K (when the sample is in the AF state) is characterized by
three prominent phonon peaks: a pronounced sharp high-energy peak (P$_1$)
at (31.0$\,\pm\,$0.2)\,meV, a less pronounced medium-energy peak (P$_2$) at
(25.5$\,\pm\,$0.3)\,meV, and a very broad, weak low-energy peak (P$_3$) at
(16.3$\,\pm\,$0.5) meV. Upon heating from 59\,K to 305\,K, where the sample
is still in the AF state, the overall shape of the VDOS essentially
remains the same (Fig.\ \ref{fig:ExpVDOS}b), however, we observe a slight red shift
(about 1\,\%, averaged over the spectrum) due to the
effect of lattice thermal expansion. When the Fe$_{48}$Rh$_{52}$ sample is heated to 416\,K, the VDOS (Fig.\ \ref{fig:ExpVDOS}c) distinctly changes in two aspects: (i)
the height of the high-energy VDOS peak P$_1$ is drastically reduced to the height of P$_2$ while its width increases;
(ii) the broad (but weak)
low-energy feature P$_3$, centered at $\sim$16 meV, becomes remarkably
narrower. In fact, the largest relative change in the broad feature P$_3$
upon heating to 416\,K occurs at $\sim$10 meV, implying a reduction of an
apparent shoulder (P$_4$) at (10.0$\,\pm\,$1.5) meV that exists at 305\,K and 59\,K, but not at 416\,K. Since $T = 416$\,K is above the transition
temperature of $\sim$380\,K, the sample Fe$_{48}$Rh$_{52}$ is in the FM
state. This transition is clearly seen in the differences between the VDOS in Figs.~\ref{fig:ExpVDOS}(a,b) and Fig.~\ref{fig:ExpVDOS}(c).
Apparently, the phonon spectrum of B2-ordered FeRh
depends on the type of magnetic ordering (AF or FM). Since the
transition occurs isostructurally, the drastic magnetism-dependent modification of the VDOS
observed in Fig.\ \ref{fig:ExpVDOS} is an atomistic manifestation of strong
magnetoelastic (or spin-phonon) interaction in the magnetocaloric
FeRh compound. In this respect the FeRh alloy behaves similarly to the
magnetocaloric ordered La(Fe,Si)$_{13}$ compound, for which also a distinct
magnetic-order-dependent modification of the Fe-projected VDOS has
been discovered by NRIXS~\cite{gruner:15}.

\begin{figure}[tbh]
\begin{center}
\includegraphics[angle=0, width=0.95\columnwidth]{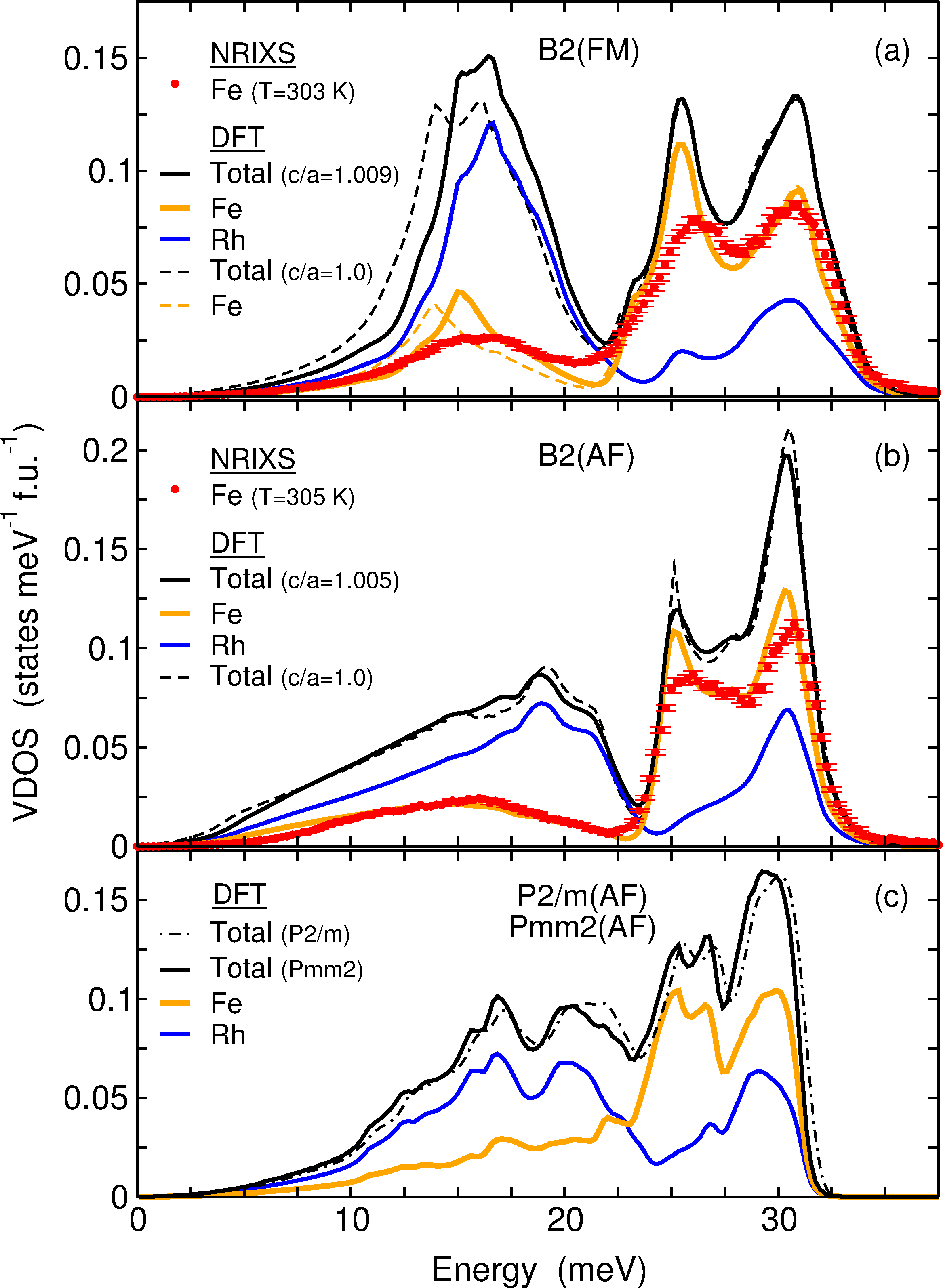}
\end{center}
\caption{
  Experimental Fe-projected VDOS (red circles with error bars) of FM Fe$_{51}$Rh$_{49}$
  measured by NRIXS at 303\,K (a) and of
  AF Fe$_{48}$Rh$_{52}$ 
  measured by NRIXS at 305\,K (b)
  compared with the element-resolved (orange for Fe, blue for Rh) and total (black)
  VDOS calculated from DFT at the respective experimental lattice constants (thick solid lines).
  This leads to slighly tetragonal cells
  ($c/a=1.009$ for FM and $c/a=1.005$ for AF according to the expitaxial strain), which we compare
  to cubic systems ($c/a=1.0$) calculated at
  the respective atomic volume (dashed lines). This reveals
  a significant shift of the low energy peak at  \unit[$\sim15$]{meV} on $c/a$.
  The lower panel (c) shows the total VDOS of the monoclinic $P$2/$m$
  AF ground state (dash-dotted black line)
  together with the total and
  element resolved VDOS (thick black lines, same colors as above) for the
  orthorhombic $Pmm$2. Both structures yield very similar results which
  differ significantly from the measured VDOS in subfigures (a) and (b).
  All VDOS curves are specified in
  states per degree of freedom, meV and f.u.\ (formula unit or element) of stoichiometric FeRh.
}\label{fig:ExpTheo}
\end{figure}

In Fig.\ \ref{fig:ExpTheo}a we present the experimental Fe-projected (partial) VDOS
of the FM sample Fe$_{51}$Rh$_{49}$ (sample code FeRh02) obtained by NRIXS at 303\,K.
For better comparison, in Fig.\ \ref{fig:ExpTheo}b we display again the
experimental partial VDOS of the AF sample Fe$_{48}$Rh$_{52}$ at 305\,K, taken
from Fig.\ \ref{fig:ExpVDOS}b. The VDOS of the FM and AF are
clearly distinguishable. In fact, the features of the VDOS
of the FM state, as shown in Fig.\ \ref{fig:ExpTheo}a, are remarkably similar to
those of the FM state above the AF-to-FM transition, (Fig.\ \ref{fig:ExpVDOS}c), as
described above. There is only a small red-shift ($\sim$2.2\,\% averaged
over the phonon spectrum) between the experimental FM VDOS in Fig.\ \ref{fig:ExpTheo}a
(at 303\,K) and Fig.~\ref{fig:ExpVDOS}c (at 416\,K) due to lattice thermal
expansion, otherwise both VDOS have nearly the same shape.
The peak positions of the FM VDOS at 63\,K
are located at P$_1 = (31.4 \pm 0.2)\,$meV , P$_2 = (26.2 \pm 0.3)\,$meV and
P$_3 = (16.6 \pm 0.5)\,$meV,
and thus agree with the corresponding peak positions of the AF VDOS
at low temperature (59\,K) (Fig.\ \ref{fig:ExpVDOS}a).
Besides the difference in peak heights, P$_1$ and P$_2$ becomes broader in the FM state than in the AF state (Fig.\ \ref{fig:ExpTheo}), the reverse is observed for
P$_3$ and P$_4$.

It is important to compare the prominent features in our experimental
Fe-projected VDOS of AF Fe$_{48}$Rh$_{52}$ with those of the [111] phonon
dispersion obtained almost 40 years ago by inelastic neutron
scattering on B2-ordered bulk FeRh at RT \cite{castets:77}.
From the extrema (minima
or maxima) observed in the [111] dispersion curve one would expect to
find van-Hove-type anomalies in the corresponding VDOS. From the
phonon dispersion in Fig.\ 3 of Ref.\ \cite{castets:77}, we find a maximum at $\sim$31 meV
and a minimum at $\sim$26 meV for the optical phonon modes, and a maximum
at $\sim$24 meV and a minimum at $\sim$17 meV for the longitudinal acoustic
phonon mode ( we used the conversion 1\,THz$\,=\,$4.132\,meV). The observed
peak positions in the partial VDOS of our AF Fe$_{48}$Rh$_{52}$ sample
(Fig.\ \ref{fig:ExpVDOS}a) are 31.0\,meV (P$_1$), 25.5\,meV (P$_2$) and 16.3\,meV (P$_3$). A
comparison shows that the latter peaks agree reasonably well with the
position of extrema in the [111] phonon dispersion. This allows us to
assign our peak P$_1$ and P$_2$ to the transverse and longitudinal optical
mode, respectively, and P$_3$ to the longitudinal acoustic mode. The
VDOS-shoulder at $\sim 10\,$meV has no counterpart in the [111] dispersion
curve.

We find excellent agreement between the position of peaks in the
experimental (NRIXS) VDOS and the positions of van Hove singularities
expected from the  phonon dispersion  relations in Fig.\ \ref{fig:Phonons}
computed at the respective equilibrium volume.
For the AF phase,
we expect computed van Hove singularities at
\unit[$10.2 \pm 2.9$]{meV}, \unit[$17.5 \pm 1.8$]{meV}, \unit[$24.1 \pm 0.5$]{meV}, and \unit[$29.2 \pm 1.1$]{meV},
as compared with our experimental
Fe-projected VDOS peaks at
\unit[$10.0 \pm 1.5$]{meV} (P$_4$), \unit[$16.3 \pm 0.5$]{meV} (P$_3$), \unit[$25.5 \pm 0.3$]{meV} (P$_2$), and \unit[$31.0 \pm 0.2$]{meV} (P$_1$). 
The agreement is remarkable in view of the
fact that no adjustable parameters were employed.
The same is valid for the FM phase: We expect computed van Hove singularities at 
\unit[$14.1 \pm 1.9$]{meV}, \unit[$24.9 \pm 1.8$]{meV}, and \unit[$30.4 \pm 0.6$]{meV},
in comparison with
\unit[$16.6 \pm 0.5$]{meV} (P$_3$), \unit[$26.2 \pm 0.3$]{meV} (P$_2$), and \unit[$31.4 \pm 0.2$]{meV} (P$_1$)
in the experiment. The remaining disagreement is remedied if the VDOS
is compared to a calculation
carried out at the experimental lattice parameters
as shown in Fig.\ \ref{fig:ExpTheo}.
This applies in particular to Peak P$_3$ in the B2(FM) phase,
for which we obtain \unit[$14.1 \pm 1.9$]{meV} from Fig.\ \ref{fig:Phonons} vs.\ $(16.6 \pm 0.5)\,$meV
in the experiment.
In the latter case, the epitaxial condition implies a slight tetragonal distortion. This indeed causes a significant shift (1.5\,meV) of P$_3$ to higher energies, which is seen by comparing the cubic FM VDOS (dashed yellow lines) to the tetragonal distorted one (solid yellow lines) at the same volume per atom (Fig.\ \ref{fig:ExpTheo}a).
From the qualitative difference between
the experimental AF-VDOS in Fig.\ \ref{fig:ExpTheo}b and the
calculated results for the predicted new AF ground state
structures shown in Fig.\ \ref{fig:ExpTheo}c, we infer that no significant fraction of
P2/m and Pmm2 structures is present in the experimental samples down to at least 60\,K. This becomes
evident from the deep minimum between P$_2$ and P$_3$ in the B2(AF) phase at 23\,meV,
which is fully reproduced by the experiment, while the computational VDOS of the the monoclinic and orthorhombic structures is larger by one order of magnitude.

Recently, anomalous structural behavior across the metamagnetic transition of Fe$_{49}$Rh$_{51}$ thin films on MgO(001) has been observed by temperature-dependent extended X-ray absorption fine structure (EXAFS) studies~\cite{wakisaka:15}. The authors extracted the $T$-dependence of the EXAFS dynamical Debye-Waller factor (or mean-square relative atomic displacement $C_2 = \langle(r - \langle r \rangle)^2\rangle)$  of Fe-Rh, Fe-Fe and Rh-Rh from the Fe and Rh K-edge EXAFS signals. Anomalous thermal behavior near the AF to FM transition (including a thermal hysteresis) was observed in particular in $C_2$ for Fe-Fe and Rh-Rh, but less so for Rh-Fe and Fe-Rh. As NRIXS measures the $T$-dependence of the total  Fe mean-square displacement, $\langle x^2 \rangle$, it is interesting to compare our NRIXS $\langle x^2 \rangle$ values with the  EXAFS $C_2$ data in Fig.~5 of Ref.~\onlinecite{wakisaka:15}. 
We observe a tendency for our $\langle x^2 \rangle$ data to be slightly lower for the FM phase than for the AF state at corresponding temperatures. Compared to the EXAFS $C_2$ values (as plotted in Fig.~5 of Ref.~\onlinecite{wakisaka:15}) our Fe $\langle x^2 \rangle$ NRIXS data are close to the $C_2$ values for Fe-Rh vibrations, but distinctly different to the $C_2$ values for Fe-Fe vibrations. This observation suggests that the dominant contribution to the NRIXS $\langle x^2 \rangle$ originates from nearest-neighbor Fe-Rh vibrational modes.
On the other hand, the larger differences in the Fe-Fe and Rh-Rh Debye-Waller factors compared to the mixed case reported by EXAFS indicates that in the AF phase vibrations are enhanced in the planes formed by either Fe or Rh alone. This could be a manifestation of the soft mode at $X$ as described in Sec.~\ref{sub:Results}, which is present in the AF but not in the FM phase (see also the inset of Fig.~\ref{fig:frozen_phonon}).

From the experimental VDOS, g(E), we obtain the Fe-projected lattice
entropy $S^{\rm vib}_{\rm Fe}(M_{\rm exp},V_{\rm exp})$ and lattice specific heat
$C^{\rm vib}_{\rm Fe}(M_{\rm exp},V_{\rm exp})$
corresponding to constant volume $V_{\rm exp}$ and magnetization $M_{\rm exp}$
at the respective measurement temperature $T_{\rm exp}$~\cite{fultz:10}.
The experimental difference
$\Delta C^{\rm vib}({\rm Fe}) = C^{\rm vib}_{\rm FM}({\rm Fe}) - C^{\rm vib}_{\rm AF}({\rm Fe})$
between the FM and AF
states in the measured temperature range is very small.
Near 60\,K, the
difference is found to be
$\Delta C^{\rm vib}({\rm Fe}) = [0.701(2) - 0.645(2)]
\,k_{\rm B}$/${\rm Fe} = + 0.056(3)\,k_{\rm B}$/Fe,
i.e., $C^{\rm vib}_{\rm FM}({\rm Fe}) > C^{\rm vib}_{\rm AF}({\rm Fe})$.
At room temperature, we find
$\Delta C^{\rm vib}({\rm Fe}) = [2.757(8) - 2.765(7)]\,k_{\rm B}$/${\rm Fe} = - 0.008(10)\,
k_{\rm B}/$Fe, being zero within error bars.
Also the difference obtained from DFT,
$\Delta C^{\rm vib}({\rm Fe})= - 0.003\,k_{\rm B}/$Fe, is very small.

Likewise, the difference
$\Delta S^{\rm vib}({\rm Fe})=S^{\rm vib}_{\rm FM}(\mathrm{Fe})-S^{\rm vib}_{\rm AF}(\mathrm{Fe})$
between the FM and AF states in the measured temperature range is
found to be small and changes sign.
Near 60\,K we obtain
$\Delta S^{\rm vib}({\rm Fe}) = [0.261(1) - 0.254(1)]\,k_{\rm B}/\mbox{Fe} = 0.0070(14)\,k_{\rm B}/\mbox{Fe}$, with
$S^{\rm vib}_{\rm FM}({\rm Fe}) > S^{\rm vib}_{\rm AF}({\rm Fe})$.
At RT, for example,
we find that
$\Delta S^{\rm vib}({\rm Fe}) = [3.345(9) - 3.410(8)]\,k_{\rm B}/\mbox{Fe} = 
- 0.065(12)\,k_{\rm B}/$Fe,
with $S^{\rm vib}_{\rm FM}({\rm Fe}) < S^{\rm vib}_{\rm AF}({\rm Fe})$, which is consistent with the increased mean square displacements in the AF phase observed by both NRIXS here and EXAFS in Ref.~\onlinecite{wakisaka:15}.

From our calculations, we obtain a somewhat larger absolute value of
$\Delta S^{\rm vib}({\rm Fe})=-0.114\,k_{\rm B}/{\rm Fe}$ at RT.
For the joint contribution of both elements to the
entropy change our calculations yield
$\Delta S^{\rm vib}=-0.210\,k_{\rm B}$/f.u., which increases significantly to
$\Delta S^{\rm vib}=-0.058\,k_{\rm B}$/f.u.,
if we calculate it for the
corresponding cubic systems at the same atomic volume.\footnote{
  The unit $k_{\rm B}$/Fe multiplied by the factor 8.480 (for Fe$_{51}$Rh$_{49}$,
  95\,\%  enriched in $^{57}$Fe) yields the unit J\,K$^{-1}$(mole Fe$_{51}$Rh$_{49}$)$^{-1}$.
  The unit $k_{\rm B}$/Fe multiplied by the factor 7.982 (for Fe$_{48}$Rh$_{52}$, 95\,\% enriched in $^{57}$Fe) yields the unit J\,K$^{-1}$(mole Fe$_{48}$Rh$_{52}$)$^{-1}$.
  The unit $k_{\rm B}$/Fe multiplied by
  the factor 53.356 (for Fe$_{51}$Rh$_{49}$, 95\,\% enriched in $^{57}$Fe) yields the unit J\,K$^{-1}$kg$^{-1}$,
  and the unit $k_{\rm B}$/Fe multiplied or by the factor 49.369 (for Fe$_{48}$Rh$_{52}$,
  95\,\% enriched in $^{57}$Fe) yields the unit J\,K$^{-1}$kg$^{-1}$.
  1\,mol of $^{57}$Fe-enriched Fe$_{51}$Rh$_{49}$
  (Fe$_{48}$Rh$_{52}$) corresponds to 158.93\,g (161.69\,g).
  For comparison, the conversion factor between $k_{\rm B}$/f.u.\ (1\,f.u.\,=\,2 atoms) or
  $k_{\rm B}$/element and J\,K$^{-1}$kg$^{-1}$ of natural stoichiometric FeRh amounts to 52.371. 
}
Since the VDOS of the AF phase remains essentially unchanged by the tetragonal distortion,
this difference is predominantely caused by the shift of the peak P$_3$ in the FM phase.
It is not advisable to compare these values directly with the total entropy change from bulk
experiments, since the different composition of the films affect the lattice parameters
and thus decrease the volume change at the phase transition, which, according to Gr\"uneisen theory, has a
considerable impact on the entropy change. We may, however, compare the entropy
change with and without tetragonal distortion, since the volume of each phase is
kept constant. From the values given above for $\Delta S^{\rm vib}$
we might expect an increase of the transition temperature by
$\Delta T\approx T\,(\Delta S^{\rm tetra}-\Delta S^{\rm cubic})/C_p\approx 8\,$K
from the impact of a tetragonal distortion of less than 1\,\% on
the vibrational entropy of the FM phase.
Uniaxial strain conditions might also evolve in bulk systems from the large volumetric
stress during the transformation. This possibly contributes to the rather large
hysteresis associated with the metamagnetic transition.
In turn, it has been shown recently, that
multi-stimuli cycles\cite{liu:12} combining the magnetocaloric transition
with biaxial compressive strain can effectively decrease
hysteresis losses.\cite{liu:16} 

\section{Thermodynamic stability from quasiharmonic calculations}
\label{sec:thermo}
To quantify the relevance of the lattice and electronic
degrees of freedom for the thermodynamic stability of the B2(FM), the B2(AF),
and the new hypothetical low temperature phase, we calculate the free energies from first-principles
within the quasiharmonic approximation~\cite{fultz:10,hickel:12,wang:13}
and derived $V(T)$, $C_p(T)$ and $S(T)$,
which
can be directly compared with experiments.
We approximate the monoclinic low temperature structure by an orthorhombic
model (point group $Pmm$2). This saves significant computation time since the unit cell lacks
the small monoclinic distortion, but still exhibits
a stable phonon dispersion. It also has a closely related VDOS to the $P$2/$m$ ground state and is extremely close in energy (\unit[+1.6]{meV/f.u.}).

Minimization of the free energy $F=E-T\,S$ with respect to
the volume $V$ at a given temperature $T$
yields the Gibbs free energy $G$ at zero pressure.
Magnetic contributions to the Gibbs free energy are not included here.
The corresponding
thermal expansion arises from the volume dependence of the
vibrational and electronic DOS.
We obtain a similar temperature dependence in all three phases with a
linear thermal expansion coefficient at room temperature of
$1.3\times 10^{-5}$\,K$^{-1}$ for B2(AF) and
$1.1\times 10^{-5}$\,K$^{-1}$ for B2(FM) and $Pmm$2, which corresponds reasonably well to the value of 
$0.95\times 10^{-5}$\,K$^{-1}$ obtained by Ibarra and Algabarel
for the AF phase of FeRh~\cite{ibarra:94}. The calculated volume change between
B2(AF) and B2(FM) of 1.5\,\% at $T=350\,$K slightly exceeds the
experimental reports of around 1\,\%~\cite{zakharov:64,levitin:66}.

\begin{figure}[tbh]
\begin{center}
\includegraphics[angle=0, width=0.95\columnwidth]{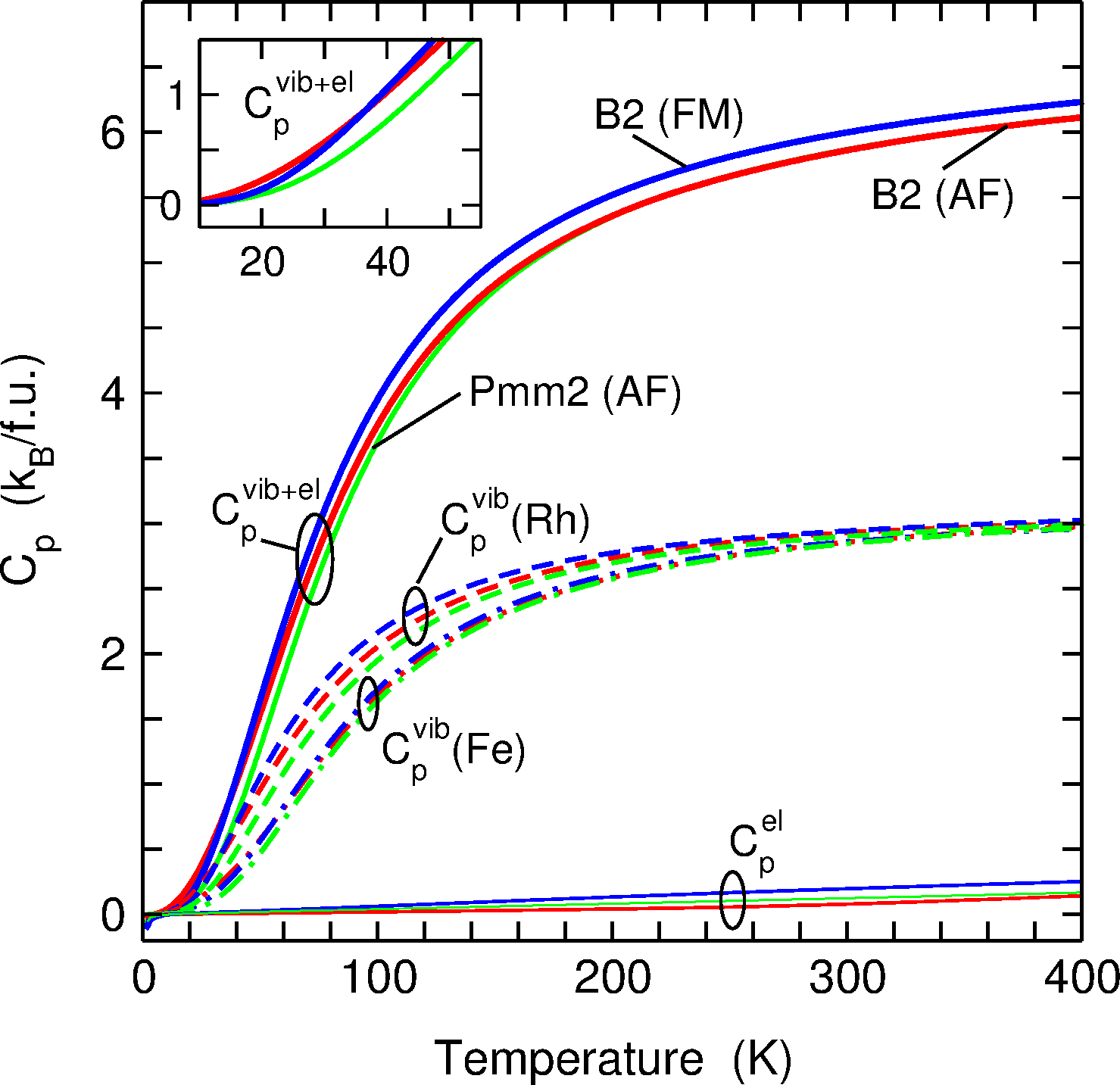}
\end{center}
\caption{Lattice and electronic specific heat $C_p$ at zero pressure
  for the B2(AF) (red), the B2(FM) (blue), and the orthorhombic $Pmm$2(AF) (green) FeRh structures.  
  The plot shows the sum of electronic and lattice specific heat (thick solid lines)
  and the element-resolved contributions to the vibrational degrees of freedom
  (dashed lines for Rh and dash-dotted lines for Fe).
  The contributions from Rh are larger than from Fe.
  The excess in $C_p$ of the FM state at room temperature and above is entirely related
  to the electronic contribution (thin solid lines).
  Below $T=200\,$K there is also a notable contribution from the
  lattice degrees of freedom corresponding to Rh. The inset shows the crossover in
  $C_p^{\rm vib+el}$ of B2(AF) and B2(FM) at low temperatures.}
\label{fig:Cp}
\end{figure}
Knowledge of $V(T)$ yields access to $S(T)$ and the specific heat at constant pressure
$C^{\rm vib+el}_p=T\,(\partial S/\partial T)_p$ arising
from the electronic and lattice degrees of freedom.
This quantity is shown in Fig.\ \ref{fig:Cp}, further decomposed into the electronic and
element specific vibrational contributions, $C^{\rm el}_p$ and $C^{\rm vib}_p$, respectively.
We see, that from $T=40\,$K upwards the FM phase exhibits a significantly larger
$C^{\rm vib+el}_p$ than the other two phases. Below $T=200\,$K this is caused by the
larger lattice specific heat of Rh, $C^{\rm vib}_p($Rh$)$.
Since the motion of the lighter element Fe is, as expected, represented by
the higher phonon frequencies, we find $C^{\rm vib}_p($Fe$)<C^{\rm vib}_p($Rh$)$ for all $T$.
The contribution of Fe is approximately the same for all three phases.
Above room temperature, where the vibrational specific heat
at constant $V$ approaches the Dulong-Petit limit,
the difference between the phases in $C^{\rm vib+el}_p$
is dominated by the difference between the
electronic contributions, which is largest for the FM phase according
to the larger electronic DOS at $E_{\rm Fermi}$.
This corroborates that the electronic degrees of freedom
deliver an important contribution
to the thermodynamic stability at elevated temperatures,
as was suggested earlier
based on experimental work\cite{tu:69}
and confirmed later from DFT
calculations~\cite{szajek:92,szajek:94,gu:05,deak:14,polesya:16}.
Below  $T=40\,$K, the lattice specific heat of the B2(AF) phase exceeds the
contribution of the B2(FM) phase (inset of Fig.\ \ref{fig:Cp}).
 The inversion at low temperatures was predicted
in the early phenomenological analysis of Ricodeau and Melville~\cite{ricodeau:72}
and later observed
in the thin film experiments of Cooke and coworkers~\cite{cooke:12}.
Such a crossover can be inferred from the  linear coefficient of the specific heat $\gamma$
which is significantly larger for the FM phase according to the larger electronic density
of states (cf.\ Fig.\ \ref{fig:DOS}).
From a fit to our electronic entropy data below 100\,K, we obtain
$\gamma_{\rm B2(AF)}=3.84\times10^{-5}\,k_{\rm B}{\rm K}^{-1}{\rm f.u.}^{-1}=2.01\,$mJ\,kg$^{-1}$K$^{-2}$,
$\gamma_{\rm B2(FM)}=1.17\times10^{-4}\,k_{\rm B}{\rm K}^{-1}{\rm f.u.}^{-1}=6.10\,$mJ\,kg$^{-1}$K$^{-2}$ and
$\gamma_{Pmm{\rm 2(AF)}}=8.36\times10^{-5}\,k_{\rm B}{\rm K}^{-1}{\rm f.u.}^{-1}=4.38\,$mJ\,kg$^{-1}$K$^{-2}$,
for the 
B2(AF), the B2(FM) and the $Pmm$2 configurations, respectively.
These values are in good agreement with previous theoretical
and experimental estimates~\cite{tu:69,ivarsson:71,fogarassy:72,cooke:12},
but too small to account for the magnitude of the effect
shown in the inset of Fig.\ \ref{fig:Cp}, which we
rather relate to the excitation
of the low-lying soft phonon branches of the B2(AF) phase.

\begin{figure}[tbh]
\begin{center}
\includegraphics[angle=0, width=0.95\columnwidth]{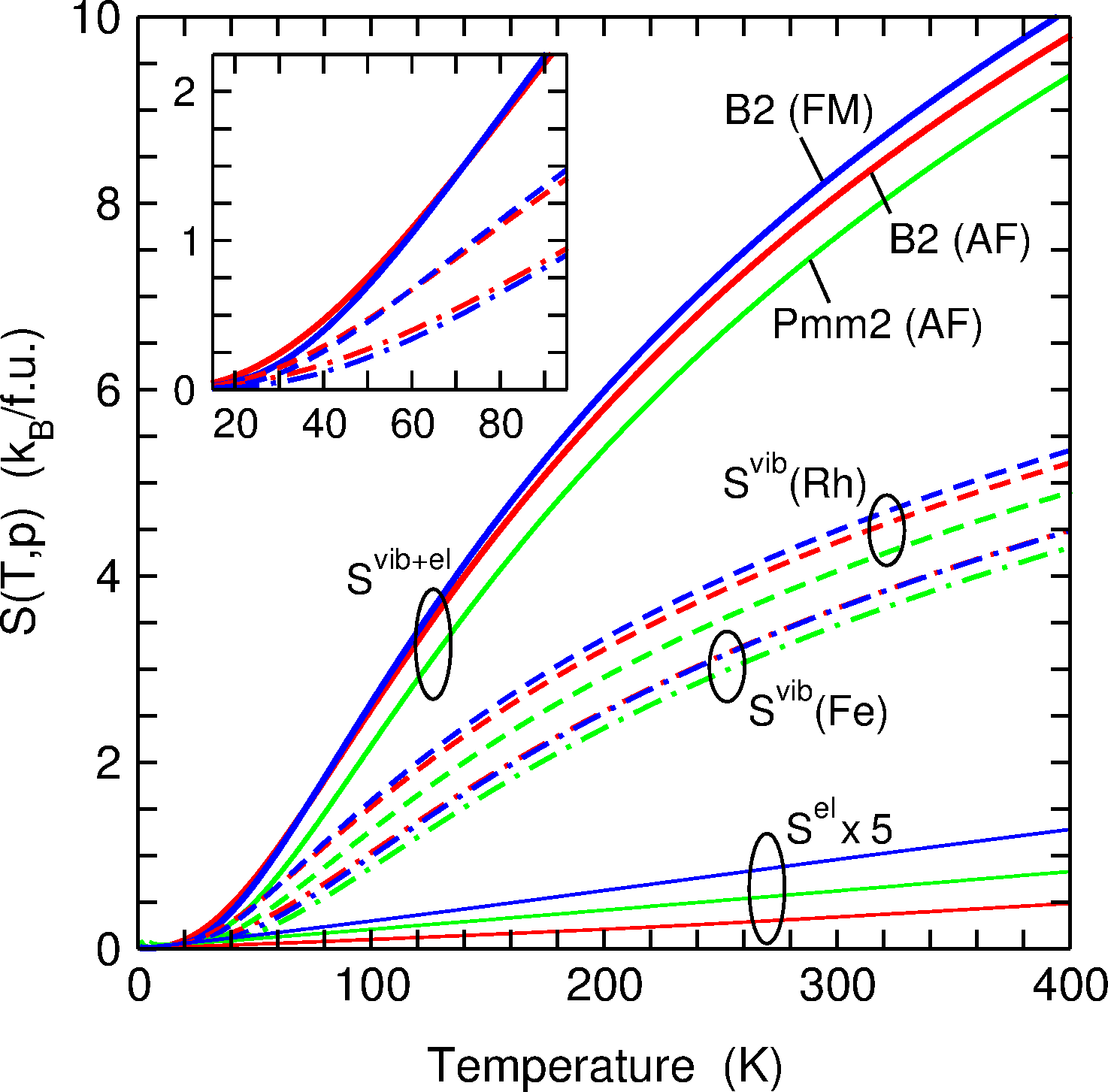}
\end{center}
\caption{
Electronic and element-resolved contributions from the lattice degrees of freedom to the
entropy $S(T,p)$ at zero pressure for B2-FeRh (AF and FM) and the $Pmm$2 structure.
Colors and line patterns correspond to Fig.\ \protect{\ref{fig:Cp}}.
Around room temperature the sum of the lattice and electronic entropy $S^{\rm vib+el}$ of the
FM phase exceeds the entropy of the AF B2-phase.
This difference is mainly related to the electronic entropy $S^{\rm el}$
(thin solid lines, enlarged by a factor of 5) and
the contribution from the
lattice entropy associated with Rh (dahes lines), originating from the excess specific heat of
the FM phase around 100\,K visible in Fig.\ \protect\ref{fig:Cp}.
The inset diplays the low temperature behavior of the total and lattice entropy.
The vibrational entropy of Fe nearly coincides for B2(AF) and B2(FM).
}\label{fig:Entropy}
\end{figure}
The larger $C^{\rm vib+el}_p$ of B2(FM) in a wide temperature range corresponds to a larger entropy $S^{\rm vib+el}$ for  $T>70\,$K.
Keeping in mind the discussion of $C_p$,
this originates from the larger vibrational contribution of Rh in combination
with the electronic entropy, which steadily increases with temperature.
At low temperatures, $S^{\rm vib+el}$ is dominated by the low lying phononic modes
in the B2(AF) phase and exceeds the entropy in the FM phase below 70\,K,
as shown in the inset of Fig.\ \ref{fig:Entropy}.

At $T=350$\,K, which corresponds to the metamagnetic transition, we obtain a
considerable difference in entropy of
$\Delta S^{\rm vib+el}=S^{\rm vib+el}_{\rm B2(FM)}-S^{\rm vib+el}_{\rm B2(AF)}=0.268\,k_{\rm B}$/f.u.$\,=\,$11.7\,J\,kg$^{-1}$\,K$^{-1}$, which is close
to the experimentally reported values of the total entropy change
obtained in field-, pressure- and temperature-induced transitions,
ranging from 12\,J\,kg$^{-1}$\,K$^{-1}$ to
19\,J\,kg$^{-1}$\,K$^{-1}$~\cite{zakharov:64,kouvel:66,richardson:73,ponomarev:72,annaorazov:92,stern-taulats:14,chirkova:16}.
Our $\Delta S^{\rm vib+el}$ is a sum of nearly equal parts of
the electronic entropy and the lattice contribution of Rh.
The vibrational degrees of freedom of Fe apparently do not contribute to the
difference in entropy between B2(FM) and B2(AF) in the relevant temperature range.
Their contribution might even be slightly negative, since the quasiharmonic calculations
overestimate the volume change at the phase transition by 0.5\,\%.
Our computational results are thus not conflicting
with our experimental measurements, which clearly show
$S^{\rm vib}_{\rm AF}($Fe$)>S^{\rm vib}_{\rm FM}($Fe$)$ at RT, as this can be traced back to
the tetragonal distortion of the films in combination with a smaller volume
difference (cf.\ Sec.\ \ref{sec:experiment}),
which is about half of the reported value for bulk FeRh. 

Comparing $\Delta S^{\rm vib+el}$
with the earlier theoretical estimates for  $\Delta S^{\rm mag}$ ranging between
6\,J\,kg$^{-1}$\,K$^{-1}$, 8\,J\,kg$^{-1}$\,K$^{-1}$ (Refs.\ \onlinecite{gruner:03} and
\onlinecite{derlet:12}, from empirical
spin-model calculations) and 14\,J\,kg$^{-1}$\,K$^{-1}$
(Ref.\ \onlinecite{gu:05}, from the magnon density of states),
we conclude that lattice, electronic, and magnetic degrees of freedom contribute
in roughly equal magnitude to the metamagnetic
transition. 

\begin{figure}[tbh]
\begin{center}
\includegraphics[angle=0, width=0.9\columnwidth]{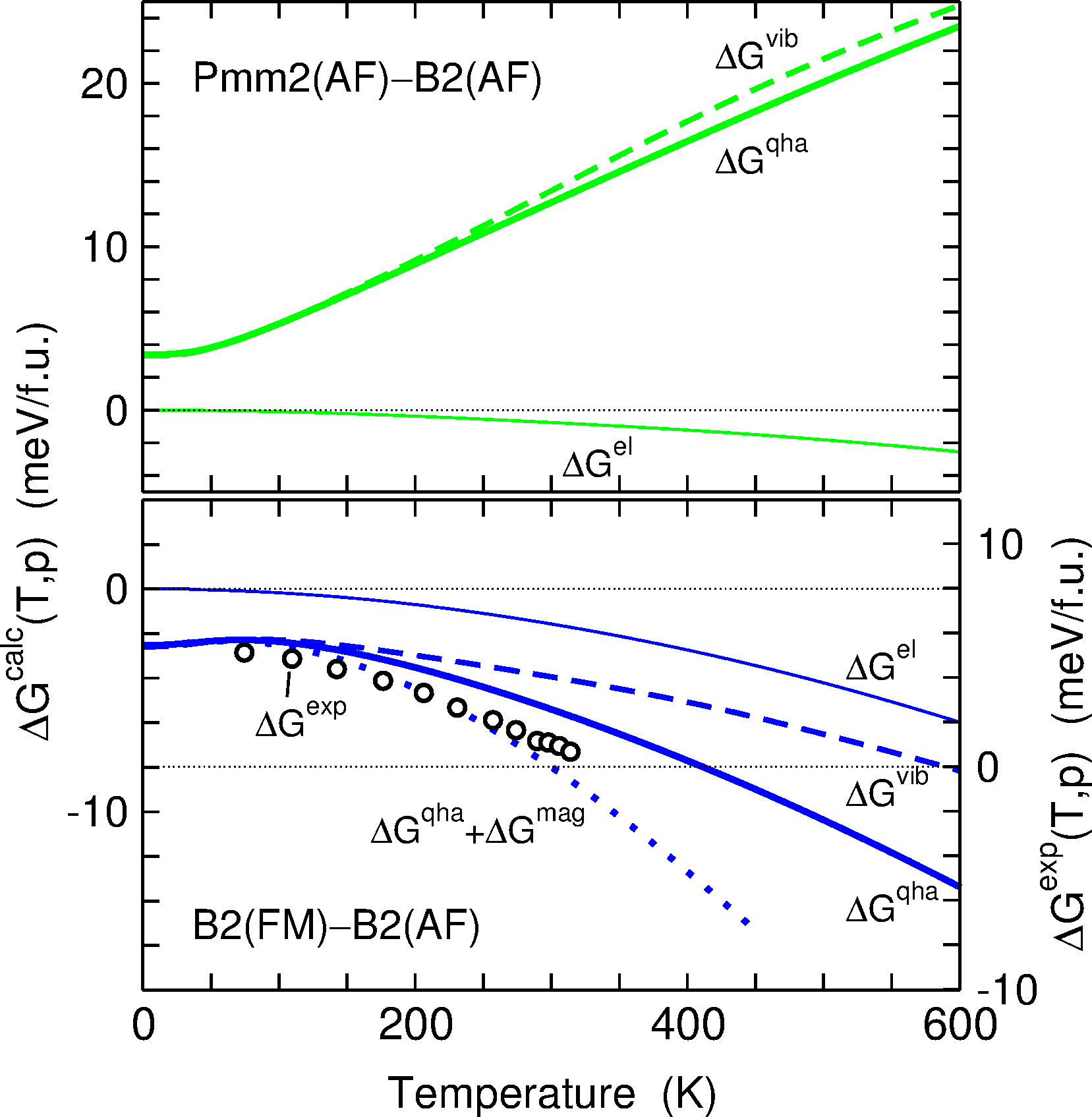}
\end{center}
\caption{Left axis: Contributions to the calculated Gibbs free energy difference $\Delta G^{\rm calc}$ 
  of the orthorhombic $Pmm$2 structure (upper panel)
  and the cubic B2(FM) phase (lower panel)
  relative to the B2(AF) phase as a function  of temperature.
  Thick lines refer to the quasiharmonic contribution $\Delta G^{\rm qha}$ according to
  Eq.\ (\protect\ref{eq:fqha}), thin solid lines to the electronic part $\Delta G^{\rm el}$
  and dashed lines to the pure vibrational contribution $\Delta G^{\rm vib}$.
  The dotted line in the lower panel denotes the sum of $\Delta G^{\rm qha}$ and the estimate of the
  magnetic Gibbs free energy difference $\Delta G^{\rm mag}$ calculated by Gu
  and Antropov\protect\cite{gu:05} from the magnon density of states.
  The open circles show the experimental Gibbs free energy differences
  $\Delta G^{\rm exp}$ and refer to the right axis.
  The scales of both axes are matched by an appropriate offset
  $\Delta G^{\rm exp}(0)-\Delta G^{\rm vib}(0)=7.97\,$meV/f.u.
  such that experimental and theoretical datasets may be directly compared.
}\label{fig:Free}
\end{figure}
The observation that the $Pmm$2 structure has the lowest specific heat
for $T<250$\,K is reflected in a significantly lower entropy compared to the B2 phases
in the entire temperature range (cf.\ Fig.\ \ref{fig:Entropy}). In relation to the B2(AF) phase the $Pmm$2(AF) phase,
at $T=350$\,K, has a rather large
$\Delta S^{\rm vib+el}=-0.434\,k_{\rm B}/{\rm f.u.}=-22.7\,$J\,kg$^{-1}$\,K$^{-1}$.
Once again, this difference is dominated by the lattice contribution of Rh
($\Delta S^{\rm vib}($Rh$)=-0.313\,k_{\rm B}$).

Fig.~\ref{fig:Free} shows the different contributions to the Gibbs free energy difference 
of the $Pmm$2(AF) and B2(FM) phases with respect to the B2(AF) phase.
In addition to the electronic and lattice part, we also show the quasiharmonic Gibbs free energy
$G^{\rm qha}(T)$, which includes all contributions,
except for the magnetic part which has not been calculated in this work and the
DFT ground state total energy $E(V(0))$, which turns out to be particularly
sensitive to methodological details:
\begin{equation}
G^{\rm qha}(T) = G^{\rm vib}(T) + G^{\rm el}(T) + E(V(T)) - E(V(0))
\label{eq:fqha}
\end{equation}
According to the large $\Delta S^{\rm vib}$,
the difference in the lattice contributions $\Delta G^{\rm vib}(T)$ between 
$Pmm$2(AF) and B2(AF) is steeply increasing with temperature.
Thus, the hypothetical low temperature structure is
suppressed with respect to the B2(AF) phase by lattice entropy,
providing a further reason for the absence of experimental indications
for a new ground state,
including our NRIXS $^{57}$Fe-VDOS in Fig.\ \ref{fig:ExpVDOS}a.

Concerning the isostructural metamagnetic transition, we observe a clear preference
for the FM phase,
which at ambient conditions, still arises in large parts from the difference in the
zero-point energies and the free energy of the electronic system. The magnitude of
$\Delta G^{\rm el}$ at room temperature corresponds well to the estimate
of D\'eak et al.~\cite{deak:14},
which takes into account the changes to the electronic structure around $E_{\rm Fermi}$
at finite temperatures through magnetic excitation in
the framework of the disordered local moment (DLM) approach.
Combining $\Delta G^{\rm qha}$ with the ground state energy difference
$\Delta E^{\rm DFT}$ from Table~\ref{tab:Energies},
this quantity turns out to be almost one order of magnitude too small
to account for the experimental phase transition temperature. This mismatch
does not disappear if we take into account the magnetic Gibbs free energy difference
$\Delta G^{\rm mag}$ calculated
by Gu and Antropov~\cite{gu:05} from the magnon dispersion
relations obtained with DFT calculations.
Keeping in mind the good agreement of the magnitude of $\Delta S^{\rm vib+el}$
with the experimental entropy change,
it is clear that the ground state energy differences between B2(AF) and B2(FM)
from our semi-local DFT calculations,
grossly overestimate the $T=0$ free energy
difference between the two phases.
This applies as well to the vast majority of DFT investigations although
a recent comparison of the electronic 
density of states obtained from DFT calculations and HAXPES measurements\cite{gray:12}
confirms a proper description of the electronic structure within the standard
semi-local GGA used in our
approach.
\footnote{
However, we note that there exist a few exceptions reporting a
significantly smaller energy difference, as the aforementioned work of
Gu and Antropov,\protect\cite{gu:05}
which is based on the
the less common Langreth-Mehl functional\protect\cite{langreth:81}
and likewise a recent GGA+U approach\protect\cite{takahashi:16}
accounting for additional correlation on Fe-d-sites.
}\nocite{langreth:81,takahashi:16}

Also the early experiment of Ponomarev~\cite{ponomarev:72}
yields a much smaller estimate for the Gibbs free energy difference at $T=0$ of
$\Delta G^{\rm exp}(0)=3.23\times 10^{-10}\,{\rm J/kg}=5.38\,$meV/f.u.\ for a
Fe$_{0.96}$Rh$_{1.04}$ alloy, which was confirmed in its magnitude from the specific
heat measurements of Cooke and coworkers~\cite{cooke:12}.
In contrast to $\Delta E^{\rm DFT}$, $\Delta G^{\rm exp}(0)$
already includes the zero point energies of the phonons of both phases as well as
a possible zero point contribution of the antiferromagnetic magnons
predicted from spin-wave theory~\cite{yosida:96,anderson:52,kubo:52}. As pointed out recently
by Polesya et al.\ by means of DLM calculations, the energy difference between
the FM and AF structure depends decisively on the magnetic order
and changes its sign when the magnetic order parameter drops below a critical value of
0.8~\cite{polesya:16}. On the other hand,
in their thermodynamic analysis of magnetic and electronic free energies, which
includes spin disorder but neglects lattice contributions,
D\'eak et al.~\cite{deak:14} did not encounter a transition for a volume below
30.08\,\AA$^3$/f.u., which exceeds the experimental volume range.

The data points from Ref.\ \onlinecite{ponomarev:72} coincide well with the sum of
$\Delta G^{\rm qha}(T)$ and the magnetic contribution $\Delta G^{\rm mag}(T)$, as calculated
by Gu and Antropov from the magnon density of states~\cite{gu:05}.
Combining $\Delta G^{\rm qha}(T)$ with $\Delta G^{\rm exp}(0)$ taken
from Ref.\ \onlinecite{ponomarev:72}, we obtain a transition temperature of
$T_{\rm M}^{\rm qha}=410\,$K. From the sum
$\Delta G^{\rm qha}(T)+\Delta G^{\rm mag}(T)$ a value of $T_{\rm M}^{\rm qha+mag}=300\,$K is found, which should be compared to
$T_{\rm M}^{\rm exp}=331\,$K, as reported by Ponomarev.
Keeping in mind that we neglect anharmonic contributions and cross-coupling
between the magnetic, electronic and vibrational degrees of freedom, we can rate this
an excellent agreement, which corroborates our present decomposition of the thermodynamic
quantities.

\section{Conclusions}
\label{sec:Conclusions}
With our combined  ab-initio and experimental approach we provide
a comprehensive survey on the lattice dynamical properties of the AF and FM
phases of B2 ordered FeRh and their relation to the metamagnetic transition.
Our experimental NRIXS investigation of the partial Fe VDOS in the FM and AF
phase is the first of its kind for FeRh and allows
an independent experimental assessment of the element-resolved vibrational
contributions to specific heat and entropy. 
The comparison with the element-resolved VDOS obtained from first principles
calculations yields a very good agreement for both phases. For the FM phase, the
agreement is  substantially improved
with respect to a Rh-dominated peak around 15\,meV by considering
the tiny tetragonal distortion of the thin film in the calculations. This reveals an unexpected strong sensitivity of the FM phonons
on uniaxial strain, which is not present at all in the AF phase.
Thus, strain conditions must be considered explicitly
in the comparison of the thermodynamic properties of the bulk material
and epitaxial thin films.
In particular, a Debye model
fitted to bulk elastic parameters as used in Ref.\ \onlinecite{cooke:12}
does not allow a sufficiently accurate
decomposition of the specific heat and entropy change measured in thin films.
The combination with ab-initio work as presented in this and other
recent studies~\cite{deak:14,polesya:16,zverev:16}
could in turn lead to a precise estimate for the magnetic specific heat.

In the AF phase, we encounter soft phonon branches along [110]
which become imaginary in a small fraction of reciprocal space around the $X$-point,
which is in agreement with another recent investigation.\cite{aschauer:16}
The instability acts as a precursor for a competing monoclinic AF
structure with lower energy.
A possible origin of this instability are symmetry related changes in the
electronic structure, which however, are most prominent far below the Fermi
energy. Accordingly, the Fermi surface of B2-ordered FeRh in the AF phase is
comparably small and shows no signs of apparent nesting.
This is in contrast to the case of L2$_1$ Heusler compounds like Ni$_2$MnGa,
which are similar in structure
and exhibit also a pronounced phonon instability along [110] as martensitic precursor.

For B2 FeRh the energy gain on the calculated barrier free transformation pathway is only \unit[0.2]{meV/f.u.}, so even small thermal fluctuations may suppress the transition to the monoclinic phase, although the fully distorted structure would be clearly favored in energy compared to the B2 phase. This could explain why we do not find any trace of the new phase down to a temperature of $59\,$K in our NRIXS VDOS measurements. This is in accordance with our first-principles estimate of the vibrational and electronic
Gibbs free energy in the quasiharmonic approximation, which also suggests a strong entropic suppression of the new phase with increasing temperature. 

On the other hand, the FM B2 structure is favored by lattice and electronic entropy.
Together with the magnetic free energy calculated from the magnon
dispersion relations taken from Ref.~\onlinecite{gu:05} we obtain an excellent agreement with the early
measurement of Ponomarev,\cite{ponomarev:72}
which is up to now the only experimental benchmark for the free energy.
From the close agreement between theory and experiment
for specific heat, entropy and Gibbs free energy,
we conclude that our decomposition provides a realistic estimate of the impact
of the different degrees of freedom on the metamagnetic transition:
We propose a cooperative and essentially equal contribution of the magnetic
and the combined electronic and vibrational contributions of the Rh-atoms, whereas
the lattice contribution associated with Fe is small.
The availability of detailed information on the thermodynamic contributions of
the lattice and the electrons opens a new route to benchmark the different magnetic models
and --- once specific heat, entropy, and free energy data are made available ---
may settle the long-standing dispute about the origin of the metamagnetic transition.

\begin{acknowledgments}
The authors would like to thank Sergii Khmelevskyi, Martijn Marsman
and Sebastian F\"ahler for fruitful discussions. We would also like to thank Ulrich von H\"orsten 
(Duisburg-Essen) for his outstanding technical assistence. W. Keune is grateful to S. D. Bader (Argonne) for enlightening discussions.
M. Wolloch, P. Mohn, J. Redinger, and
D. Suess acknowledge the support by the Austrian Science Fund (FWF)
[SFB ViCoM F4109-N28 and F4112-N28]. M. E. Gruner, O. Gutfleisch,
P. Entel and H. Wende acknowledge the support by the Deutsche
Forschungsgemeinschaft (DFG) within the priority program SPP 1599.
This work was supported by the DFG  SPP 1681 (WE2623/7-1), FOR 1509 (WE2623/13-2) and by Stiftung Mercator (MERCUR).
The authors also appreciate the
ample support of computer resources by the Vienna Scientific Cluster
(VSC) and the use of the Cray XT6/m supercomputer of Center for
Computational Sciences and Simulation (CCSS) at University of Duisburg-Essen.
This research used resources of the Advanced Photon Source, a U.S. Department of Energy (DOE)  Office of Science User Facility operated for the DOE Office of Science by Argonne National Laboratory under Contract No. DE-AC02-06CH11357.
Figs.~\ref{fig:Cells}, \ref{fig:frozen_phonon}, and~\ref{fig:monoclinic_cell} in this paper were created with help of the
VESTA code~\cite{vesta:11}. The FINDSYM utility of the ISOTROPY Software
Suite~\cite{findsym:05} was used for structure identification. 
\end{acknowledgments}

\end{document}